\documentclass{amsproc}
\usepackage{psfig}
\usepackage{amsmath}
\title{Perturbation Theory and Numerical Modeling \\
of Quantum Logic Operations \\
with a Large Number of Qubits}

\author{G. P. Berman$^1$, G. D. Doolen$^1$, D. I. Kamenev$^1$
\address{$^1$Theoretical Division, T-13, and CNLS,
 Los Alamos National Laboratory,
Los Alamos, NM 87545}}

\author{\ \\ G. V. L\'opez$^2$
\address{$^2$Departamento de F\'isica, Universidad de Guadalajara,
Corregidora 500, S.R. 44420, Guadalajara, Jalisco, M\'exico}}

\author{V. I. Tsifrinovich$^3$
\address{$^3$IDS Department, Polytechnic University,
Six Metrotech Center, Brooklyn NY 11201}}

\thanks{This work was supported by the Department of Energy under
contract W-7405-ENG-36, by the National Security Agency (NSA), and by the
Advanced Research and Development Activity (ARDA)}


\date{July 06, 2001.}

\begin{document}
\begin{abstract}
The perturbation theory is developed based on small parameters
which naturally appear in solid state quantum computation.
We report the simulations of the dynamics of quantum logic
operations with a large number of qubits (up to 1000). A nuclear spin
chain is considered in which selective excitations of spins are provided
by having a uniform gradient of the external magnetic field. Quantum
logic operations are utilized by applying resonant electromagnetic pulses.
The spins interact with their nearest neighbors. We simulate the creation
of the long-distance entanglement between remote qubits in the spin chain.
Our method enables us to minimize unwanted non-resonant effects in a
controlled way. The method we use cannot simulate complicated quantum
logic (a quantum computer is required to do this), but it can be useful
to test the experimental performance of simple quantum logic operations.
We show that: (a) the probability distribution of unwanted states has
a ``band'' structure, (b) the directions of spins in typical unwanted
states are highly correlated, and (c) many of the unwanted states are
high-energy states of a  quantum computer (a spin chain). Our approach
can be applied to simple quantum logic gates and fragments of quantum
algorithms involving a large number of qubits.
\end{abstract}

\maketitle

\section*{Introduction}
Recently much progress has been made in single particle technologies.
These technologies allow one to manipulate a single electron, a single
atom, and a single ion. A distinguishing feature of all these technologies
is that they are ``quantum''; quantum effects are crucial for the
preparation of the initial state, for performing useful operations, and for
reading out the final state. One of the future quantum technologies is
quantum computation. In a quantum computer, the information is loaded in
a register of quantum bits -- ``qubits''. A qubit is a quantum object
(generalized spin 1/2) which can occupy two quantum states,
$|0\rangle$ and $|1\rangle$, and an arbitrary superposition of these
states: $\Psi=C_0|0\rangle+C_1|1\rangle$. (The only constraint is:
$|C_0|^2+|C_1|^2=1$.) A quantum computer is remarkably efficient in
executing newly invented quantum algorithms (the quantum Fourier transform,
Shor's algorithm for prime factorization, Grover's algorithm for data base
searching, and others). Using these algorithms, quantum computing promises
to solve problems that are intractable on digital computers. The main
advantage of quantum computation is the rapid, parallel execution of
quantum logic operations. One of the promising directions in quantum
computation is solid state quantum computation. (See the review
\cite{ber1}.) In solid state computers, a qubit can be represented by a
single nuclear spin 1/2 \cite{kane,ber2}, or a single electron spin
1/2 \cite{loss,yab}, a single Cooper pair \cite{nak1,nak2}. This type
of quantum computers is quite different from quantum computers in which
a qubit is represented by an ensemble of spins 1/2 \cite{ger,cory,yam}.

Crucial mathematical problems must be solved in order to understand the
dynamical aspects of quantum computation.
 One of these problems is the creation of the dynamical theory of quantum
computation -- the main subject of our paper. The processes of the creation
of quantum data bases, the storage and searching of quantum information,
the implementation of quantum logic gates, and all the steps involved in
quantum computation are dynamical processes. When qubits representing a
register in a quantum computer are in superpositional states, they are
not eigenstates of the Hamiltonian describing the quantum computer. These
superpositions are time-dependent. Understanding their dynamics is very
important. To design a working quantum computer, simulations of quantum
logic operations and fragments of quantum computation are essential. The
results of these simulations will enable engineers to optimize and test
quantum computers.

There are two main obstacles to perform useful quantum logic operations
with large number of qubits on a digital (classical) computer: (1) the
related Hilbert space is extremely large, $D_N=2^N$, where $N$ is the
number of qubits (spins), and (2) even if the initial state of a quantum
computer does not involve many basic states (eigenstates), the number of
excited eigenstates can rapidly grow during the process of  performing
quantum logic operations.  Generally both of these obstacles exist.
At the same time, many useful quantum operations can be
simulated on a digital computer even if the number of qubits is quite
large, say 1000. How one can do this? One way is to create (and use)
a {\it perturbation theory of quantum computation}. To do this, one
should consider a quantum computer as a many-particle quantum system
and introduce small parameters. Usually, in {\it all} physical problems
there exist small parameters which allow one to simplify the problem to
find approximate solutions. The objective in this approach is to build
a solution in a controlled way.

As an example, it is useful to consider an electron in a hydrogen atom
interacting with a laser field. The electron has an infinite (formally)
number of discrete levels plus a continuous part of its energy spectrum.
So, a single electron in a Hydrogen atom has a Hilbert space larger than
{\it all} finite qubit  quantum computers. Assume that the electron is
populated initially on some energy level(s). The action of the laser field
on the electron leads in many cases to a regular and controlled dynamics
of the electron. In particular, if the amplitude of the laser field is
small enough, the electron interacting with this laser field can be
considered as a two-level quantum system interacting with an external
time-periodic field. Why is the electron excited to only few energy
levels? This happens because the existence of small parameters makes the
probability of unwanted events (excitation of the electron to most energy
levels including the continuum) negligibly small. The existence of small
parameters allows one to use perturbation approaches when calculating the
quantum dynamics of the electron. This strategy can be used to create
a perturbation theory of quantum computation.

In this paper we present our results, based on the existence
of small parameters, for simulating quantum logic operations with a large
number (up to 1000) of qubits. Our perturbation approach does
not provide a substitution
for computing quantum algorithms on a quantum computer. One will need
quantum computer to do the required calculations. Our method will allow
one to simulate the required benchmarks, quantum logic test operations,
and fragments of quantum algorithms.  Our method will enable engineers to
optimize a working quantum computer.

In section 1, we formulate our model
and the Hamiltonian of a quantum computer based on a one-dimensional
nuclear spin chain. Spins interact through
nearest-neighbor Ising interactions. All spins are placed in a
magnetic field with the uniform
gradient. This field gradient enables the selective excitation of
spins. Quantum logic operations are provided by applying the required
resonant electromagnetic pulses. In section 2, we present the dynamical
equations of motion in the interaction representation. As well,
we discuss an alternative description based on the transformation
to the rotating frame. In the latter case the effective
Hamiltonian is independent of time and solution of the problem
can be obtained using the eigenstates of this Hamiltonian.
In section 3,
we derive simplified dynamical equations taking into account only
resonant and near-resonant transitions (and neglecting non-resonant
transitions).
We introduce the small parameters which characterize the probability of
generation of unwanted states in the result of near-resonant transitions.
We describe a $2\pi k$-method which allows us
to minimize the influence of unwanted near-resonant effects which can
destroy quantum logic operations.
In section 4, we describe a quantum
Control-Not gate for remote qubits in a quantum computer with a large
number of qubits. Analytical solution for this gate is presented in
section 5, for the $2\pi k$ condition. In section 6, we present small
parameters of the problem, in explicit form.
The equation for
the total probability of generation of unwanted states, including
the states generated in the result of near-resonant and non-resonant
transition, is derived in section 7.
In section 8, we present results of numerical simulations
of the quantum Control-Not gate for remote qubits in quantum computers
with 200 and 1000 qubits when the probability of the near-resonant
transitions is relatively large (when the conditions of $2\pi k$-method
are not satisfied).
We show that: (a) the probability distribution
of unwanted states has a ``band'' structure, (b) the directions of spins
in typical unwanted states are highly correlated, and (c) many of the
unwanted states are high-energy states of the  quantum computer (a spin
chain).
The total probability of error (including
the states generated in the result of near-resonant and non-resonant
transitions) is computed numerically in section 9 for small ($N=10$) and
large ($N=1000$) number of qubits. A range of parameters is found in which
the probability of error does not exceed a definite threshold.
We test our perturbative approach using exact numerical solution
when the number of qubits is small ($N=10$).
In section 10, we show how the problem of quantum computation
(even with large number of qubits) can be formulated classically in
terms of interacting one-dimensional oscillators. In the conclusion,
we summarize our results.

\section{Formulation of the model}
The mathematical model of a quantum computer used in this paper is based
on a one-dimensional Ising nuclear-spin system -- a chain of $N$ identical
nuclear spins (qubits). Application of Ising spin systems for quantum
computations was first suggested in Ref.\cite{ber3}. Today, these  systems
are used in liquid nuclear magnetic resonance (NMR) quantum computation
with small number of qubits \cite{chuang}.
The register (a 1D chain of $N$ identical nuclear spins)
is placed in a magnetic field,
\begin{equation}
\label{1}
{\vec B}(t,\,z)=
\left(b^{(n)}_\perp\cos\left[\nu^{(n)} t+\varphi^{(n)}\right],
-b^{(n)}_\perp\sin\left[\nu^{(n)} t+\varphi^{(n)}\right], B^z(z)\right),
\end{equation}
where $t^{(n)}\le t\le t^{(n+1)}$ and $n=1,...,M$.
In (1), $B^z(z)$ is a slightly non-uniform magnetic
field oriented in the positive $z$-direction. The quantities
$b^{(n)}_\perp>0$, $\nu^{(n)}>0$ and $\varphi^{(n)}$
are, respectively, the amplitude, the frequency, and the initial phase
of the circular polarized (in the $x-y$ plane) magnetic field. This
magnetic field has the form of rectangular pulses of the length (time
duration) $\tau^{(n)}=t^{(n+1)}-t^{(n)}$. The total number of pulses
which is required to perform a given quantum computation (protocol) is
$M$. Schematically, our quantum computer is shown in Fig. 1.

\begin{figure}[t]
\centerline{\psfig{file=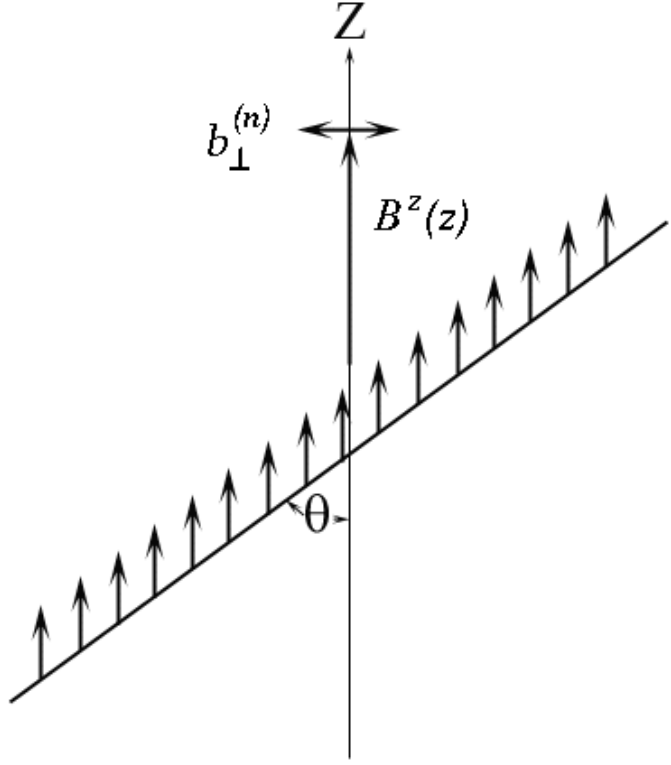,width=12cm,height=12cm}}
\vspace{-10mm}
\caption{Nuclear spin quantum computer (the ground state of nuclear spins).
$B^z(z)$ is the slightly non-uniform magnetic field; $b^{(n)}_{\perp}$
is the amplitude of radio-frequency field of the $n$-th {\it rf} pulse.
The chain of spins makes an angle $\theta$ with the direction of the
magnetic field $B^z(z)$.}
\label{fig:1}
\end{figure}

\subsection{The quantum computer Hamiltonian}
The quantum Hamiltonian which describes our quantum computer has the form,
\begin{equation}
\label{2}
\hat{H}=\hat {H}_0+\hat {V}=-\sum_{k=0}^{N-1}\omega_k \hat
I^z_k-2J\sum_{k=0}^{N-2}\hat I^z_k\hat I^z_{k+1}+\hat {V}.
\end{equation}
(We set $\hbar=1$.) The operator $\hat {V}$ describes the
interaction of spins with pulses of the {\it rf} field,
\begin{equation}
\label{3}
\hat{V}=\sum_{n=1}^M\hat{V}^{(n)},
\end{equation}
and $\hat{V}^{(n)}$ describes the interaction of spins
with $n$-th pulse of the {\it rf} field,
\begin{equation}
\label{4}
\hat {V}^{(n)}=-\Theta^{(n)}(t){{\Omega^{(n)}}\over{2}}\sum_{k=0}^{N-1}
\Bigg\{\hat I^-_k\exp\left[-i\left(\nu^{(n)}t+\varphi^{(n)}\right)\right]+
\end{equation}
$$
\hat I^+_k\exp\left[i\left(\nu^{(n)}t+\varphi^{(n)}\right)\right]\Bigg\},
$$
where $\Theta^{(n)}(t)$ equals 1 only during the $n$th pulse.
The operators in (\ref{2})-(\ref{4}) have the following explicit form in
the $z$-representation (the representation in which the operators
$\hat I^z_k$ are diagonal),
\begin{equation}
\label{5}
\hat I^z_k=
{{1}\over{2}}|0_k\rangle\langle 0_k|-{{1}\over{2}}|1_k\rangle\langle 1_k|,
\end{equation}
$$
\hat I^+_k=\hat I^x_k+i\hat I^y_k=|0_k\rangle\langle 1_k|,~ 
\hat I^-_k= \hat I^x_k-i\hat I^y_k= |1_k\rangle\langle 0_k|.
$$
We shall use the Dirac notation for the complete set of eigenstates
(the stationary states) of the quantum computer described by the
Hamiltonian $\hat{H}_0$ in (\ref{2}). The eigenstates of the spin chain can
be described as a combination of $2^N$ individual states of nuclear spins,
\begin{equation}
\label{6}
|00...00\rangle,~|00...01\rangle,..., |11...11\rangle,
\end{equation}
where the state $|0\rangle$ corresponds to the direction of a nuclear
spin along the direction of the magnetic field, $B^z(z)$, and the state
$|1\rangle$ corresponds to the opposite direction. The subscript ``$k$''
is omitted.

In (\ref{2})-(\ref{4}), $\omega_k$ is the Larmor frequency of the $k$-th
spin (neglecting interactions between spins), $\omega_k=\gamma B^z(z_k)$,
and $\gamma$ is the nuclear gyromagnetic ratio. (For example, for a proton
in the field $B^z(z_k)=10$T, one has the nuclear magnetic resonance (NMR)
frequency $f_k=\omega_k/2\pi\approx 430$MHz.) We assume, for definiteness,
that the gradient of the magnetic field is positive,
$\partial B^z(z)/\partial z>0$. Suppose that the frequency difference of
two neighboring spins is, $\delta f=\delta\omega/2\pi\approx 10$kHz,
where $\delta\omega=\omega_{k+1}-\omega_{k}$.
Thus, if the frequency of the edge spin is $430$MHz, the frequency of the
spin at the other end of the chain of $1000$ qubits
is $\approx 440$MHz. Then, the value
of $B^z(z)$ increases by $\Delta B^z=0.23$T along the spin chain. Taking
the distance between the neighboring spins, $a\approx 2\AA$, we obtain
the characteristic value for the gradient of the magnetic field,
$\partial B^z(z)/\partial z\approx 0.23/1000 a\cos\theta$, where
$\theta$ is the angle between the direction of the chain and the
$z$-axis (Fig. 1). Below we will take $\cos\theta=1/\sqrt{3}$.
(This allows one to suppress the dipole interaction between spins in
the eigenstates of the Hamiltonian $\hat{H}_0$.) Thus, the gradient
of the magnetic field is
$\partial B^z(z)/\partial z\approx 2\times 10^6$T/m.
The quantity $\Omega^{(n)}=\gamma b^{(n)}_{\perp}$ is
the Rabi frequency of the $n$-th pulse.

\section{Quantum dynamics of the computer}
We discuss below the dynamics of the spin chain described by the
Hamiltonian (\ref{2})-(\ref{4}). The quantum state of the quantum
computer is described by the wave function $\Psi(t)$. The dynamics
of this function
is given by the solution of the Schr\"odinger equation,
\begin{equation}
\label{14}
i\dot\Psi(t)=\hat {H}\Psi(t),
\end{equation}
where dot means derivative over time. This linear equation and
an initial condition, $\Psi(0)$, define the state of the system
at the time $t$. In this section we describe two different
approaches which allow to solve Eq. (\ref{14}).

\subsection{Equations of motion in the interaction representation}
In order to compute the dynamics
it is convenient to go over to interaction representation.
The wave function, $\Psi_{int}(t)$, in the
interaction representation, is connected to the wave function, $\Psi(t)$,
in the laboratory system of coordinates by the transformation,
\begin{equation}
\label{11}
\Psi(t)=\hat U_0\Psi_{int}(t),~\hat U_0=\exp(-i\hat{H}_0t),
\end{equation}
where, $\hat{H}_0$ is defined in (\ref{2}).
We choose the eigenfunctions, $|p\rangle$, and the eigenvalues, $E_p$,
of the Hamiltonian $\hat{H}_0$ as the basis states and
expand the wave function $\Psi_{int}(t)$ in a series,
\begin{equation}
\label{Cp}
\Psi_{int}(t)=\sum_{p=0}^{2^N-1}C_p(t)|p\rangle,
\end{equation}
where the states $|p\rangle$ satisfy the equation
\begin{equation}
\label{12}
\hat{H}_0|p\rangle=E_p|p\rangle,~p=1,...,2^N.
\end{equation}
The wave function in the laboratory system of coordinates has the form,
\begin{equation}
\label{13}
\Psi(t)=\sum_pC_p(t)|p\rangle\exp(-iE_pt),
\end{equation}
Using the Schr\"odinger equation (\ref{14}) for the wave function,
$\Psi(t)$, we obtain the equations of motion for the amplitudes $C_p(t)$,
\begin{equation}
\label{15}
i\dot C_p=\sum_{m=0}^{2^N-1}
C_m{V}^{(n)}_{pm}\exp\left[i(E_{p}-E_m)t+ir_{pm}
\left(\nu^{(n)}t+\varphi^{(n)}\right)\right],
\end{equation}
which represent the wave function in the interaction representation.
Here $r_{pm}=\mp 1$ for $E_p>E_m$ and $E_p<E_m$, respectively,
${V}^{(n)}_{pm}=-\Omega^{(n)}/2$ for the states $|p\rangle$
and $|m\rangle$ which are connected by a single-spin transition,
and ${V}^{(n)}_{pm}=0$ for all other states.

\subsection{The dynamics in the rotating frame}
Sometimes, to calculate the quantum dynamics generated by the
Hamiltonian (\ref{2}) during the action of the $n$-th pulse,
$t^{(n)}\le t\le t^{(n+1)}$, it is convenient to make a transformation
to the system of coordinates which rotates with the frequency
$\nu^{(n)}$ of the magnetic field, $\vec{b_{\perp}}$.
To do this, one can use the unitary transformation \cite{ber4},
\begin{equation}
\label{7}
\hat U^{(n)}=\exp\Bigg[i\left(\nu^{(n)}t+\varphi^{(n)}\right)
\sum_{k=0}^{N-1}\hat I^z_k\Bigg].
\end{equation}
In the rotating system of coordinates the new wave function,
$\Psi_{rot}(t)$ and the new Hamiltonian, ${\mathcal H}^{(n)}$ have
the form,
\begin{equation}
\label{8}
\Psi_{rot}(t)=\hat U^{(n)+}\Psi(t),
\end{equation}
\begin{equation}
\label{9}
\hat {\mathcal H}^{(n)}=
\hat U^{(n)+}\hat{H}^{(n)}\hat U^{(n)}=
-\sum_{k=0}^{N-1}\Bigg[\left(\omega_k-\nu^{(n)}\right)
\hat I^z_k+\Omega^{(n)}\hat I^x_k+2J\hat I^z_k\hat I^z_{k+1}\Bigg],
\end{equation}
where the term $\hat I^z_{N-1}\hat I^z_{N}$ must be excluded.
In the rotating system of coordinates the effective
magnetic field which acts on the $k$-th spin during the action of the
$n$-th pulse has the components,
\begin{equation}
\label{10}
B^{(n)}_k=\left(\Omega^{(n)},0,\omega_k-\nu^{(n)}\right)/\gamma.
\end{equation}

The advantage of the Hamiltonian (\ref{9}) is that it is time-independent.
In this case one can use the eigenstates of the Hamiltonian
$\hat{\mathcal H}^{(n)}$ to calculate the dynamics during the $n$th pulse
without solution of the system of differential equations (\ref{15}).
As before, we expand the wave function
in the rotating frame, $\Psi_{rot}(t)$, over the eigenstates of the
Hamiltonian $\hat{H}_0$,
\begin{equation}
\label{Cp1}
\Psi_{rot}(t)=\sum_{p=0}^{2^N-1}A_{p}(t)|p\rangle.
\end{equation}
The wave function in the laboratory system of coordinates is,
\begin{equation}
\label{Psi1}
\Psi(t)=
\sum_pA_{p}(t)|p\rangle\exp\left(-i\chi_{p}^{(n)}t+i\xi_{p}^{(n)}\right),
\end{equation}
where
$\chi_{p}^{(n)}=-\left(\nu^{(n)}/2\right)\sum_{k=0}^{2^N-1}\sigma_k^{p}$,
$\xi_{p}^{(n)}=\left(\varphi^{(n)}/2\right)\sum_{k=0}^{2^N-1}\sigma_k^{p}$,
$\sigma_k^{p}=-1$ if the state $|p\rangle$ contains the $k$th spin in
in the state $|1\rangle$,
and $\sigma_k^{p}=1$ if the state $|p\rangle$ contains the $k$th spin
in the state $|0\rangle$. Below we take $\varphi^{(n)}=\xi_{p}^{(n)}=0$ for
all $n$. The Schr\"odinger equation for the amplitudes
$A_p(t)$, has the form,
\begin{equation}
\label{Sch2}
i\dot A_p(t)=\left(E_p-\chi_p^{(n)}\right)A_p(t)-
\frac{\Omega^{(n)}}2\sum_{p'}A_{p'}(t),
\end{equation}
where the sum is taken over the states $|p'\rangle$
connected by a single-spin transition with the state $|p\rangle$.

Equation (\ref{Sch2}) can be written in the form
$i\dot A_p(t)={\mathcal H}^{(n)}_{pp'}A_{p'}(t)$,
where the effective Hamiltonian,
$\hat{\mathcal H}^{(n)}$, is independent of time. When the number of spins, $N$, is
not very large ($N<30$), one can diagonalize the
Hamiltonian matrix ${\mathcal H}^{(n)}_{pp'}$ and find its
(time-independent) eigenfunctions,
\begin{equation}
\label{psi}
\psi^{(n)}_q=\sum_pA_p^{q\,(n)})|p\rangle
\end{equation}
and eigenvalues, $e^{(n)}_q$. The amplitudes $A_p^{q\,(n)}$ are
the eigenfunctions of the Hamiltonian $\hat{\mathcal H}^{(n)}$
in the representation of the Hamiltonian $\hat{H}_0$.

Using the eigenstates of the
Hamiltonian $\hat{\mathcal H}^{(n)}$ one can calculate the dynamics generated
by the $n$th pulse as,
\begin{equation}
\label{dynamics}
A_p(t_{n})=A_p=
\sum_{p_0}A_{p_0}\sum_{q}A_{p_0}^{q\,(n)}A_p^{q\,(n)}
\exp\left(-ie_q^{(n)}\tau_n\right),
\end{equation}
where $t_{n}=t_{n-1}+\tau_n$, $\tau_n$ is the duration of the pulse,
$A_{p_0}=A_p(t_{n-1})$ is the amplitude before the action of the $n$th
pulse, $A_p=A_p(t_{n})$ is the amplitude after the action of the $n$th
pulse.

The amplitudes, $A_p(t)$, in the rotating frame are related to the
amplitudes, $C_p(t)$, in the interaction representation by
(see Eqs. (\ref{13}) and (\ref{Psi1})),
\begin{equation}
\label{AC}
A_p(t)=\exp(-i{\mathcal E}_p^{(n)}t)C_p(t),
\end{equation}
Where ${\mathcal E}_p^{(n)}=E_p-\chi_p^{(n)}$ are the diagonal elements
of the Hamiltonian matrix ${\mathcal H}_{pp'}^{(n)}$.
Since the {\it rf} pulses in the protocol are different, the effective
time-independent Hamiltonians, ${\mathcal H}_{pp'}^{(n)}$,
are also different for different $n$.
Hence, before each $n$th pulse we should make the transformation to the
rotating frame, $C_p(t_{n-1})\rightarrow A_p(t_{n-1})$, and after
the pulse we must return to the interaction representation
$A_p(t_{n})\rightarrow C_p(t_{n})$.

In the rotating frame the most difficult
problem is the diagonalization of the sparse
symmetric matrices (for each pulse) of the size $2^N\otimes 2^N$.
(In each row of this matrix there are only $N+1$ nonzero matrix elements.)
This approach can
be used directly (without perturbative consideration) only
for small quantum computers, with the number of qubits $N<30$.
For our purposes, to model quantum dynamics in a quantum computer with
large number of qubits ($N=1000$), we shall use a perturbation approach
based on the existence of small parameters.

\section{Resonant and near-resonant transitions}
The number of equations in (\ref{15}) grows exponentially as
$N$ increases. Thus, this system of equations cannot be solved
directly when $N$ is large. Our intention is to solve
Eqs. (\ref{15}) or (\ref{Sch2})
approximately, but in a controlled way.
This can be done if we make use of small parameters which exist for
the system. The explicit expressions for small parameters will be
given below, in sections \ref{sec:simulat} and \ref{sec:numerical}.

\subsection{Resonant and non-resonant frequencies}
In this section, we describe the main
simplification procedure which is based on the separation of resonant
and non-resonant interactions. This procedure is commonly used in studying
linear and non-linear dynamical systems \cite{bog,lich,zas}. The main
idea of this approach is the following. When the frequency of the external
field, $\nu^{(n)}$, is equal (or close) to the eigenfrequency of the
system under consideration,  $\omega_{pm}\equiv E_p-E_m$, and when a
small parameter exists (the interaction is relatively weak), one can
separate the slow (resonant) dynamics from the fast (non-resonant)
dynamics. The main contribution to the evolution of the system is
associated with resonant effects. Usually, non-resonant effects are
small, and can be considered using perturbation approaches. In quantum
computation, the problem of non-resonant effects is
more complicated because these effects can accumulate
in time creating significant errors in the process of quantum computation.
So, in quantum computation, the non-resonant effects must not only be
taken into consideration, but they must also be minimized. Below, we
describe methods which allow us to minimize the influence of non-resonant
effects.

In classical dynamical systems, the separation of resonant and non-resonant
effects is very efficient when ``action-angle'' variables are used. In this
case, the ``action'' is a slow variable and  the ``angle'' (the ``phase'')
is a fast variable. From this point of view, Eqs. (\ref{15}) are
convenient because they are written in the ``energy'' representation in
which the slow (resonant) effects and fast (non-resonant) effects can be
naturally separated.

First, we note that as the number of spins, $N$, increases, the number of
eigenstates, $|p\rangle$, increases exponentially (as $2^N$), but the
number of {\it resonant frequencies} in our system grows only linearly
in $N$. Indeed, the number of resonant frequencies is $3N-2$ because only
single-spin transitions are allowed by the operator $\hat {V}$ in
(\ref{3}). The definition of the resonant frequency is the following.
We introduce the frequency of the single-spin transition:
$\omega_{pm}\equiv E_p-E_m$, where $E_p$ and $E_m$ are two eigenstates,
$|p\rangle$ and $|m\rangle$, of the Hamiltonian $\hat{H}_0$ with
opposite orientations of only the $k$-th spin, ($k=0,...,N-1$).
For example,
\begin{equation}
\label{16}
|p\rangle=|n_{N-1}n_{N-2}...0_k...n_1n_0\rangle,~|m\rangle=
|n_{N-1}n_{N-2}...1_k...n_1n_0\rangle,
\end{equation}
where $n_k=0,1$. The resonant frequency for transition between $p$th and
$m$th states,
$\nu^{(n)}_{res}$, of the $n$-th external {\it rf} pulse is defined
as follows:
\begin{equation}
\label{17}
\nu^{(n)}_{res}=|\omega_{pm}|.
\end{equation}
The only resonant frequencies of our nuclear spin quantum computer are,
\begin{equation}
\label{18}
~~~\,\qquad\omega_k\pm J,\qquad\qquad \,\,(k=0,~or~k=N-1),
\end{equation}
$$
\omega_k,~\omega_k\pm 2J, \qquad(1\le k\le N-2),
$$
where the Larmor frequencies, $\omega_k=\gamma B^z(z_k)$.
All resonant frequencies in (\ref{18}) are assumed to be positive.
For end spins with $k=0$ and $k=N-1$, the upper and lower signs in
(\ref{18}) correspond to the states $|0\rangle$ or $|1\rangle$ of their
only neighboring spins. For inner spins with $1\le k\le N-2$ the frequency
$\omega_k$ corresponds to having nearest neighbors whose spins are in
opposite directions to each other.  The ``+'' sign corresponds to having
the nearest neighbors in their ground state. The ``-'' sign corresponds
to having the nearest neighbors in their excited  state.

As an example,
consider the resonant frequency of a single-spin transition between the
following two eigenstates,
$$
|p\rangle=|n_{N-1}n_{N-2}...1_{k-1}0_k1_{k+1}...n_1n_0\rangle,
$$
$$
|m\rangle=|n_{N-1}n_{N-2}...1_{k-1}1_k1_{k+1}...n_1n_0\rangle.
$$
These two eigenstates differ by a transition of the $k$-th spin from its
ground state, $|0_k\rangle$, to its excited state,
$|1_k\rangle $. All other spins in the quantum computer remain in their
initial states. According to (\ref{2}), the eigenvalues of the states
$|p\rangle$ and $|m\rangle$ are:
\begin{equation}
\label{20}
E_p=E_{nres}-{{\omega_k}\over{2}}+J, \qquad
E_m=E_{nres}+{{\omega_k}\over{2}}-J,
\end{equation}
where $E_{nres}$ is the total energy of all $N-1$ (except the $k$-th)
which did not participate in the transition. It follows from (\ref{20})
that in this case, the resonance frequency, $\nu^{(n)}_{res}$, of
the $n$-th pulse is,
\begin{equation}
\label{21}
\nu^{(n)}_{res}=|E_p-E_m|=\omega_k-2J.
\end{equation}
This frequency belongs to the set of the resonant frequencies introduced
in (\ref{18}).
In the following, we assume that our protocol includes only the resonant
frequencies from (18), and we will omit the subscript ``res''.

\subsection{Dynamical equations for resonant and near-resonant transitions}
In this sub-section we derive the approximate equations for resonant and
near-resonant transitions, in the system. Consider an arbitrary eigenstate
of the Hamiltonian $\hat{H}_0$,
\begin{equation}
\label{22}
|n_{N-1}n_{N-2}...n_1n_0\rangle,
\end{equation}
where, as before, the subscript indicates the position of the spin in a
chain, and $n_k=0,1$. Assume that one applies to the spin chain a resonant
{\it rf} pulse of a frequency, $\nu^{(n)}$, from (\ref{18}). Then, one has
two possibilities:\\
1) The frequency of the pulse, $\nu^{(n)}$, is the resonant frequency of
the state (\ref{22}).\\
2) The frequency, $\nu^{(n)}$, differs from the closest resonant frequency
of the state (\ref{22}) by the value $\Delta=\pm 2J$ or $\pm 4J$
(near-resonant frequency).

In the first case, one has a resonant transition. In the second case,
one has a near-resonant transition. If the small parameters exist,
$J/\delta\omega\ll 1$, and $\Omega/\delta\omega\ll 1$, we can neglect
all non-resonant transitions related to a flip of $k'$th non-resonant spin.
Below, in section \ref{sec:numerical}, we will write a rigorous
condition for $\delta\omega$ which
is required in order to neglect all non-resonant transitions.

Thus, considering the dynamics of any state (\ref{22}) under
the action of an {\it rf} pulse with the frequency, $\nu^{(n)}$,
from (\ref{17}), (\ref{18}),
we need only to take into account one transition.
This transition will be either a
resonant one or a near-resonant one.
This allows us to reduce the system of equations (\ref{15}) to the set of
only two coupled equations (two-level approximation),
\begin{equation}
\label{23}
~~~~~~i\dot C_p=-(\Omega^{(n)}/2)\exp[i(E_p-E_m-\nu^{(n)})t]C_m,
\end{equation}
$$
i\dot C_m=-(\Omega^{(n)}/2)\exp[i(E_m-E_p+\nu^{(n)})t]C_p,
$$
where $E_p>E_m$, $|p\rangle$ and $|m\rangle$ are any two eigenstates
which are connected by a single-spin transition and whose energies differ
by $\nu^{(n)}$ or $\nu^{(n)}+\Delta_{pm}^{(n)}$, where
$\Delta_{pm}^{(n)}=E_p-E_m-\nu^{(n)}$ is the frequency detuning
for the transition between the states $|p\rangle$ and $|m\rangle$.

\subsection{Solutions of the dynamical equations in two-level approximation}
The solution of Eqs. (\ref{23}) for the case when the system is
initially in the eigenstate $|m\rangle$, is,
\begin{equation}
\label{24}
C_m(t_0+\tau)=\left[\cos(\lambda_{pm}\tau/2)+
i(\Delta_{pm}/\lambda_{pm})\sin(\lambda_{pm}\tau/2)\right]
\times\exp(-i\tau\Delta_{pm}/2),
\end{equation}
$$
C_p(t_0+\tau)=i(\Omega/\lambda_{pm})\sin(\lambda_{pm}\tau/2)\times
\exp(it_0\Delta_{pm}+i\tau\Delta_{pm}/2),
$$
$$
C_m(t_0)=1,\qquad C_p(t_0)=0,
$$
where $\lambda_{pm}=(\Omega^2+\Delta_{pm}^2)^{1/2}$,
$t_0$ is the time of the beginning
of the pulse; $\tau$ is the duration of the pulse.
Here and below we omit the upper index ``$(n)$'' which indicates
the number of the {\it rf} pulse.
If the system is initially in the upper
state, $|p\rangle$, ($C_m(t_0)=0$, $C_p(t_0)=1$), the solution of
(\ref{23}) is,
\begin{equation}
\label{26}
C_p(t_0+\tau)=[\cos(\lambda_{pm}\tau/2)-
i(\Delta_{pm}/\lambda_{pm})\sin(\lambda_{pm}\tau/2)]
\times\exp(i\tau\Delta_{pm}/2),
\end{equation}
$$
C_m(t_0+\tau)=i(\Omega/\lambda_{pm})\sin(\lambda_{pm}\tau/2)\times
\exp(-it_0\Delta_{pm}-i\tau\Delta_{pm}/2),
$$
$$
C_p(t_0)=1,\qquad C_m(t_0)=0.
$$

For the resonant transition ($\Delta_{pm}=0$), the expressions
(\ref{24}) and (\ref{26}) transform into the well-known equations for
the Rabi transitions. For example, we have from (\ref{24}) for
$\Delta_{pm}=0$,
$$
C_m(t_0+\tau)=\cos(\Omega\tau/2),\qquad C_p(t_0+\tau)=i\sin(\Omega\tau/2).
$$
In particular, for $\Omega\tau=\pi$ (the so-called $\pi$-pulse),
the above expressions describe the complete transition from the
state $|m\rangle$ to the state $|p\rangle$.

For the near-resonant transitions, consider two characteristic parameters
in expressions (\ref{24}) and (\ref{26}):
$$
\epsilon^{(1)}_{pm}=\Omega/\lambda_{pm},\qquad
\epsilon^{(2)}_{pm}=\sin(\lambda_{pm}\tau/2).
$$
If either of these two parameters, $\epsilon^{(1)}_{pm}$ or
$\epsilon^{(2)}_{pm}$,
is zero, the probability of a non-resonant transition vanishes.  The second
parameter, $\epsilon^{(2)}_{pm}$, is equal zero when,
\begin{equation}
\label{27}
\lambda_{pm}\tau=2\pi k,\qquad (k=1,2,..),
\end{equation}
where $k$ is the number of revolutions of the non-resonant (average) spin
about the effective field in the rotating frame.
Eq. (\ref{27}) is the condition for the ``$2\pi k$''-method to eliminate the
near-resonant transitions. (See \cite{ber4}, Chapter 22, and
\cite{ber5,ber6}.)

\subsection{The $2\pi k$-method}
We shall present here the explicit conditions for the ``$2\pi k$'' rotation
from the state $|p\rangle$ to the state $|m\rangle$.
For a $\pi$-pulse ($\Omega\tau=\pi$), the values of $\Omega$ which satisfy
the $2\pi k$ condition are, according to Eq. (\ref{27}),
\begin{equation}
\label{28}
\Omega^{[k]}_{pm}=|\Delta_{pm}|/\sqrt{4k^2-1}=|\Delta_{pm}|/\sqrt{3},~
|\Delta_{pm}|/\sqrt{15},
~|\Delta_{pm}|/\sqrt{35},~|\Delta_{pm}|/\sqrt{63},...
\end{equation}
For a $\pi/2$-pulse, the corresponding values of $\Omega$ are,
\begin{equation}
\label{29}
\Omega^{[2k]}_{pm}=|\Delta_{pm}|/\sqrt{16k^2-1}=|\Delta_{pm}|/\sqrt{15},~
|\Delta_{pm}|/\sqrt{63},\dots
\end{equation}
(If the Rabi frequency, $\Omega$, satisfies the $2\pi k$ condition
(\ref{29}) for a $\pi/2$-pulse, it automatically satisfies the condition
(\ref{28}) for a $\pi$-pulse.)

\subsection{Resonant and near-resonant transitions in the rotating frame}
We consider here the structure of the effective
time-independent Hamiltonian matrix ${\mathcal H}_{pp'}$
in the rotating frame. We discuss the two-level approximation
in the rotating system of coordinates and demonstrate equivalence
of the solution in the rotating frame with the solution
(\ref{24}) (or (\ref{26})) in the interaction representation.

Let us discuss the structure of the matrix ${\mathcal H}_{pp'}$.
Since the spins in the chain are identical, all nonzero
non-diagonal matrix elements are the same and equal to ($-\Omega/2$).
At $\Omega\ll\delta\omega$ the absolute values of the
diagonal elements in general case
are much larger than the absolute values of the
off-diagonal elements. We explain below how
the resonance is coded in the structure of the diagonal
elements of the Hamiltonian matrix, ${\mathcal H}_{pp'}$.

Suppose that the $k$th spin in the chain has resonant or near-resonant
NMR frequency. The energy separation between $p$th and $m$th
diagonal elements of the matrix ${\mathcal H}_{pp'}$
related by the flip of resonant (or near-resonant) $k$th spin
is much less than the energy separation
between the $p$th diagonal elements and diagonal
elements related to other states, which differ from the state
$|p\rangle$ by a flip of a non-resonant $k'$th
($k'\ne k$) spin.
In this case one can neglect the interaction of the $p$th state with
all states except for the state $|m\rangle$, and the Hamiltonian matrix
${\mathcal H}_{pp'}$ breaks up into $2^N/2$,
approximately independent $2\times 2$ matrices,

\begin{equation}
\label{2x2}
\begin{pmatrix}
{\mathcal E}_{m} & V  \\
V & {\mathcal E}_{p}  ,
\end{pmatrix}
\end{equation}
where $E_{p}=E_{m}+\Delta_{pm}$,
$|\Delta_{pm}|\sim J$ or zero, and $V=-\Omega/2$ is the perturbation
amplitude.

Suppose, for example,  that $N=5$ and third spin ($k=3$) has
resonant or near-resonant frequency.
(We start enumeration from the right as shown in (\ref{16}).)
Then, the block $2\times 2$ will be organized, for example,
by the following states: $|01010\rangle$ and $|00010\rangle$;
$|01111\rangle$ and $|00111\rangle$; $|00001\rangle$ and $|01001\rangle$,
and so on. In order to find the state $|m\rangle$, which form $2\times 2$
block with a definite state $|p\rangle$, one should flip
the resonant spin of the state $|p\rangle$. In other words,
positions of $N-1$ (non-resonant) spins of these states are equivalent,
while position of the resonant spin is different.

We now obtain the solution in the two-level approximation.
The dynamics is given by Eq. (\ref{dynamics}). Since we deal only with
a single $2\times 2$ block of the matrix ${\mathcal H}_{pp'}$,
(but not with the whole matrix), the dynamics in this approximation
is generated only by the eigenstates of one block. In order to demonstrate
equivalence of two descriptions (in the rotating frame and in the
interaction representation) we will choose the same initial condition
as in (\ref{24}). Then, we will make the transformation (\ref{AC})
to the rotating frame, $C_p\rightarrow A_p$. After that we shall compute
the dynamics by Eq. (\ref{dynamics}) and return to the interaction
representation, $A_p\rightarrow C_p$, using Eq. (\ref{AC}). Eventually,
we will obtain Eq. (\ref{24}). The result is the same
in the rotating frame and in the
interaction representation. Physically these two approaches are equivalent
(exactly, but not only in the two-level approximation),
but mathematically they are different. In the interaction representation
we calculate the dynamics generated by the time-dependent Hamiltonian.
There are no stationary states in this case and one should solve
the system of differential equations. On the other hand, in the rotating
frame the effective Hamiltonian $\hat{\mathcal H}$ is
time-independent and the wave function evolves in time because it is
not the eigenfunction of $\hat{\mathcal H}$.

The eigenvalues
$e^{(0)}_q$, $e^{(0)}_Q$, and the eigenfunctions
of the $2\times 2$ matrix (\ref{2x2}) are
(we put $\Delta_{pm}=\Delta$ and $\lambda_{pm}=\lambda$),
\begin{equation}
\label{q01}
e^{(0)}_q={\mathcal E}_m+\frac\Delta 2-\frac\lambda 2, \qquad
\begin{pmatrix}
A^{q\,(0)}_m \\ A^{q\,(0)}_p
\end{pmatrix}=
{1\over\sqrt{(\lambda-\Delta)^2+\Omega^2}}
\begin{pmatrix}\Omega\\ \lambda-\Delta\end{pmatrix},
\end{equation}
\begin{equation}
\label{q02}
e^{(0)}_Q={\mathcal E}_m+\frac\Delta 2+\frac\lambda 2, \qquad
\begin{pmatrix}
A^{Q\,(0)}_m\\ A^{Q\,(0)}_p
\end{pmatrix}=
{1\over\sqrt{(\lambda-\Delta)^2+\Omega^2}}
\begin{pmatrix}
-(\lambda-\Delta)\\ \Omega
\end{pmatrix}.
\end{equation}
Suppose that before the $n$th pulse the system is in the state $|m\rangle$,
i.e. the conditions
$$
C_m(t_0)=1,\qquad C_p(t_{0})=0,
$$
are satisfied. After the transformation, (\ref{AC}), to the rotating frame
we obtain
$$
A_m(t_{0})=\exp(-i{\mathcal E}_mt_{0})C_m(t_{0})
=\exp(-i{\mathcal E}_mt_{0}),\qquad A_p(t_{0})=0.
$$
The dynamics is given by Eq. (\ref{dynamics}), which in our case
takes the form:
$$
A_m(t)=
A_m(t_{0})\left[\left(A_m^{q\,(0)}\right)^2\exp\left(-ie^{(0)}_q\tau\right)+
\left(A_m^{Q\,(0)}\right)^2\exp\left(-ie^{(0)}_Q\tau\right)\right]=
$$
$$
{\exp\left[-i[{\mathcal E}_mt-(\Delta/2)\tau]\right]\over
\Omega^2+(\lambda-\Delta)^2}\left\{\Omega^2 e^{-i\lambda\tau/2}+
(\lambda-\Delta)^2e^{i\lambda\tau/2}\right\},
$$
where $t=t_0+\tau$.
Applying the back transformation,
$$
C_m(t)=\exp(i{\mathcal E}_mt)A_m(t),
$$
and taking the real and imaginary parts of the expression
in curl brackets we obtain the first equation (\ref{24}).
For another amplitude we have,
$$
A_p(t)=A_m(t_{0})
\left[A_m^{q\,(0)}A_p^{q\,(0)}\exp\left(-ie^{(0)}_q\tau\right)+
A_m^{Q\,(0)}A_p^{Q\,(0)}\exp\left(-ie^{(0)}_Q\tau\right)\right]=
$$
$$
i{\Omega\over\lambda}
\exp\left\{-i\left[{\mathcal E}_mt-(\Delta/2)\tau\right]\right\}
\sin(\lambda\tau/2).
$$
Applying the back transformation,
$$
C_p(t)=\exp[i({\mathcal E}_m+\Delta)t]A_p(t),
$$
we obtain the second equation in (\ref{24}).

One may demonstrate the equivalence of our two approaches in a different,
more simple, way.
The dynamical equations with the $2\times 2$ Hamiltonian (\ref{2x2}) are,
$$
i\dot A_p={\mathcal E}_pA_p+VA_m,
$$
$$
i\dot A_m=VA_p+{\mathcal E}_mA_m.
$$
After the transformation (\ref{AC}) to the interaction representation
we obtain,
\begin{equation}
\label{correspondence}
~~~~~i\dot C_p=V\exp[i({\mathcal E}_p-{\mathcal E}_m)t]C_m,
\end{equation}
$$
i\dot C_m=V\exp[i({\mathcal E}_m-{\mathcal E}_p)t]C_p,
$$
where ${\mathcal E}_p-{\mathcal E}_m=E_p-E_m-\nu=\Delta_{pm}$, $V=-\Omega/2$. One
can see that Eqs. (\ref{correspondence}) are equivalent to
Eqs. (\ref{23}) with the solution given by (\ref{24}) or (\ref{26}).

\section{A quantum Control-Not gate for remote qubits}
The quantum Control-Not ($CN_{kl}$) gate is a unitary operator which
transforms the eigenstate,
\begin{equation}
\label{30}
|n_{N-1}...n_k.......n_l....n_0\rangle,
\end{equation}
into the state,
\begin{equation}
\label{31}
|n_{N-1}...n_k.......\bar n_l....n_0\rangle,
\end{equation}
where $\bar n_l=1-n_l$ if $n_k=1$; and $\bar n_l=n_l$ if $n_k=0$.
The $k$-th and $l$-th qubits are called the control and the target qubits
of the $CN_{kl}$ gate. One can also introduce a modified quantum $CN$ gate
which performs the same transformation (\ref{30}), (\ref{31})
accompanied by phase shifts which are different for different
eigenstates \cite{ber4}. It is well-known that the quantum
$CN$ gate can produce an entangled state of two qubits, which cannot be
represented as a product of the wave functions of the individual qubits.

In this section, we shall consider an implementation of the quantum $CN$
gate in the Ising spin chain with the left end spin as the control qubit
and the right end spin as the target qubit,
i.e. $CN_{N-1,0}$ for a spin chain with large number of qubits, $N$
($N=200$ or $N=1000$).
Using this quantum gate we will create entanglement between the end qubits
in the spin chain.

Assume that initially a quantum computer is in its ground state,
\begin{equation}
\label{32}
\Psi=|0...0\rangle.
\end{equation}
Then we apply a $\pi/2$-pulse with the resonant frequency,
\begin{equation}
\label{33}
\nu=\omega_{N-1}.
\end{equation}
The $\pi/2$-pulse means that the duration of this pulse is
$\tau=\pi/2\Omega$.
This pulse produces a superpositional state of
the $(N-1)$-th (left) qubit,
\begin{equation}
\label{34}
\Psi=(1/\sqrt{2})(|0...0\rangle+i|1...0\rangle).
\end{equation}
To implement a modified quantum $CN_{N-1,0}$ gate we apply to the spin
chain $M=2N-3$ $\pi$-pulses which transform
the state $|10\dots 00\rangle$ to the state
$|10\dots 01\rangle$ by the following scheme:
$
|1000\dots 0\rangle\rightarrow|1100\dots 0\rangle\rightarrow
|1110\dots 0\rangle\rightarrow
$
$
|1010\dots 0\rangle\rightarrow|1011\dots 0\rangle\rightarrow
|1001\dots 0\rangle\rightarrow
$
$
\dots\rightarrow|100\dots 11\rangle\rightarrow|100\dots 01\rangle.
$
All frequencies of our protocol are resonant for these transitions.
If we apply the same protocol to the system in the ground state,
then with large probability the system will remain in the ground state
because these pulses have the detunings from resonant transitions,
$\Delta_n=\Delta^{(n)}_{0m}\ne 0$. Here the ground state
$|0\rangle=|00\dots 00\rangle$ is related to the state $|m\rangle$ by
the flip of the $k$th spin with the near-resonant NMR frequency.
The first $\pi$-pulse has the frequency $\omega=\omega_{N-2}$. For the
second $\pi$-pulse $\omega=\omega_{N-3}$. For the third $\pi$-pulse,
$\omega=\omega_{N-2}-2J$, etc. All detunings, $\Delta_n$,
in our protocol
are the same, $\Delta_n=2J$, except for the third pulse, where
$\Delta_3=4J$.

\section{An analytic solution with application of a $2\pi k$-method}
Assume that all $M$ $\pi$-pulses satisfy the $2\pi k$ condition, with
the same value of $k$:
\begin{equation}
\label{35}
\lambda_n^{[k]}\tau_n=2\pi k, ({\rm and}~\Omega_n^{[k]}\tau_n=\pi),
\end{equation}
where $\lambda^{[k]}_n=\sqrt{\left(\Omega^{[k]}_n\right)^2+\Delta_n^2}$.
In this case, we can derive an analytic expression for the wave function,
$\Psi$, after the action of a $\pi/2$-pulse and $M$ $\pi$-pulses. This
solution has the form,
\begin{equation}
\label{36}
\Psi=C_0|00..0\rangle+C_1|10...1\rangle,
\end{equation}
where,
\begin{equation}
\label{37}
C_0={(-1)^{kM}\over\sqrt{2}}\exp(-i\pi M\sqrt{4k^2-1}/2),\qquad
C_1={(-1)^{N-1}\over\sqrt{2}},
\end{equation}

\subsection{Large $k$ asymptotics}
For $k\gg 1$, we get the same solution for odd and even $k$:
$$
C_0\approx 1/\sqrt{2}.
$$
This result is easy to understand. For a $\pi$-pulse, the Rabi frequency is,
\begin{equation}
\label{Omega_k}
\Omega_n^{[k]}=|\Delta_n|/\sqrt{4k^2-1}.
\end{equation}
Large values of $k$ correspond to small values of the parameter:
$$
\Omega_n^{[k]}/|\Delta_n|\ll 1.
$$
As $\Omega_n^{[k]}/|\Delta_n|$
approaches zero, the non-resonant pulse becomes unable to change
the quantum state.

\subsection{Small $k$ behavior}
For small values of $k$, the non-resonant pulse can change the phase
of the initial state. For example, for $k=1$ we have,
\begin{equation}
\label{38}
C_0=\exp[i\pi M(1-\sqrt{3}/2)]/\sqrt{2}.
\end{equation}
After the first $\pi$-pulse, the phase shift is approximately $24^o$.
This phase shift grows as the number of $\pi$-pulses, $M$, increases.
This increasing phase is an effect which can be easily controlled.

\section{Small parameters}
We shall now introduce the small parameters of the problem. Consider
the probability of non-resonant transition. This probability will be small
if $\varepsilon_n$ is small:
\begin{equation}
\label{39}
\varepsilon_n=(\Omega_n/\lambda_n)^2\sin^2(\lambda_n\tau_n/2)\ll 1.
\end{equation}
(It follows from (\ref{24}) that the expression for $|C_m|^2$
can be written in the form: $|C_m|^2=1-\varepsilon_n$, and
$|C_p|^2=\varepsilon_n$.)
In order to minimize the errors caused by the near-resonant
transitions, it is reasonable to keep the values of
$\varepsilon_n$ the same and small.
Since the values of the detuning are the same for all
pulses, $\Delta_n=\Delta=2J$ (except for the third
pulse, where $\Delta_3=4J$), and because
$\varepsilon_n$ depends only on the ratio $|\Omega_n/\Delta_n|$
we take the values of $\Omega_n$ to be the same,
$\Omega_n=\Omega$ ($n\ne 3$) and $\Omega_3=2\Omega$. In this case
$\varepsilon_n$ is independent of $n$ ($\varepsilon_n=\varepsilon$).
If we take into consideration the change
of phase of the generated unwanted state, this change will be small if,
\begin{equation}
\label{40}
\Omega_n/|\Delta_n|\approx \Omega_n/\lambda_n\ll 1.
\end{equation}
(For the $2\pi k$-method, this condition requires $k\gg 1$.)

Next, we will discuss the probability of the near-resonant transitions,
after the action of $M$ pulses, using small parameters $\varepsilon_n$.
Analytic expressions for probabilities $|C_0|^2$ and $|C_1|^2$ are:
\begin{equation}
\label{41}
|C_0|^2=\frac 12\prod_{n=1}^M(1-\varepsilon_n),\qquad
|C_1|^2=\frac 12.
\end{equation}
If all values of $\varepsilon_n$ are the same for all pulses,
$\varepsilon_n=\varepsilon$, then in the first non-vanishing
approximation of the perturbation theory we obtain,
\begin{equation}
\label{41a}
|C_0|^2=\frac 12(1-M\varepsilon),\qquad
|C_1|^2=\frac 12.
\end{equation}
The decrease of the probability, $|C_0|^2$, is caused by the generation
of unwanted states.
One can see that the deviation from the value, $|C_0|^2=1/2$, grows as the
number of $\pi$-pulses, $M$, increases.
It means that the small parameter of the problem is $M\varepsilon$
rather than $\varepsilon$. When $M\approx 2N$ is large, even a small
deviation from the $2\pi k$ condition can produce large
distortions from the desired wave function (\ref{36}).

\section{Non-resonant transitions}
In this section we estimate the errors caused by
non-resonant transitions and write the formula for the total
probability of errors including into consideration both near-resonant
and non-resonant transitions.
In order to estimate the probability of non-resonant transitions
it is convenient to work in the rotating frame, where the effective
Hamiltonian is independent of time.

Consider a transition between the states $|m\rangle$ and $|m'\rangle$
related by a flip of a non-resonant $k'$th spin.
The absolute value of the
difference between the $m$th and $m'$th diagonal elements of the matrix
${\mathcal H}_{pp'}$ is of order or greater than $\delta\omega$, because they
belong to different $2\times 2$ blocks~(\ref{2x2}).
Since the absolute values of
the matrix elements which connect the different blocks are small,
$|V|\ll\delta\omega$, we can use the standard perturbation
theory~\cite{Landau}. The wave function, $\psi_{q}$, in (\ref{psi})
can be written in the form,
\begin{equation}
\label{psi10}
\psi_{q}=\psi^{(0)}_{q}+
{\sum_{q'}}'{v_{qq'}\over e^{(0)}_q-e^{(0)}_{q'}}
\psi^{(0)}_{q'},
\end{equation}
where superscript ``$(n)$'' is omitted,
prime in the sum means that the term with $q'=q$ is omitted,
$\psi_{q}$ is the eigenfunction of the Hamiltonian ${\mathcal H}$,
the $q$th eigenstate is related to the $m$th diagonal
element and the $q'$th eigenstate is related to the $m'$th diagonal
element, $v_{qq'}=2V\langle\psi^{(0)}_{q}|I_{k'}^x|\psi^{(0)}_{q'}\rangle$
is the matrix element for transition between
the states $\psi^{(0)}_{q}$ and $\psi^{(0)}_{q'}$,
the sum over $q'$ takes into consideration all possible non-resonant
one-spin-flip transitions
from the state $|m\rangle$.
Because the matrix ${\mathcal H}_{pp'}$ is divided into
$2^{N-1}$ relatively independent $2\times 2$ blocks,
the energy, $e^{(0)}_q$ ($e^{(0)}_{q'}$),
and the wave function, $\psi^{(0)}_{q}$ ($\psi^{(0)}_{q'}$),
in (\ref{psi10}) are, respectively, the eigenvalue (\ref{q01}) and the
eigenfunction,
\begin{equation}
\label{ef0}
\psi^{(0)}_{q}=A^{q\,(0)}_m|m\rangle+A^{q\,(0)}_p|p\rangle,\qquad
(\psi^{(0)}_{q'}=A^{q'\,(0)}_m|m'\rangle+A^{q'\,(0)}_p|p'\rangle)
\end{equation}
of the single $2\times 2$ block (\ref{2x2}) with all other elements being
equal to zero.

When the block (\ref{2x2}) is related to the near-resonant
transition ($\Delta\sim J$) and when $\Omega\ll J$,
the eigenfunctions of this block are,
\begin{equation}
\label{qApprox}
~~~~~~~\psi^{(0)}_{q}\approx
[1-(\Omega^2/32J^2)]|m\rangle+(\Omega/4J)|p\rangle,
\end{equation}
$$
\psi^{(0)}_{Q}\approx -(\Omega/4J)|m\rangle+
[1-(\Omega^2/32J^2)]|p\rangle.
$$
On the other hand, if this block is related to the resonant transition
($\Delta=0$), we have,
\begin{equation}
\label{qApprox1}
\psi^{(0)}_{q}=(1/\sqrt 2)(|m\rangle+|p\rangle),\qquad
\psi^{(0)}_{Q}=(1/\sqrt 2)(|m\rangle-|p\rangle).
\end{equation}
The probability of the non-resonant transition from the state
$|m\rangle$ to the state $|m'\rangle$ connected
by a flip of the non-resonant $k'$th spin is,
\begin{equation}
\label{psi19}
P_{mm'}=\left|\langle m'|\psi_q\rangle\right|^2.
\end{equation}
Note, that only one term in the sum in (\ref{psi10})
contributes to the probability $P_{mm'}$, so that,
\begin{equation}
\label{psi9}
P_{mm'}\approx
\left({V\over {\mathcal E}_m-{\mathcal E}_{m'}}\right)^2\approx
\left({V\over |k-k'|\delta\omega}\right)^2,
\end{equation}
where we put $e^{(0)}_q\approx {\mathcal E}_m$,
$e^{(0)}_{q'}\approx {\mathcal E}_{m'}$, $v_{qq'}\approx V$,
$|k-k'|$ is the ``distance'' from the non-resonant $k'$th
spin to the resonant $k$th spin.

The total probability
of generation of all unwanted states by one pulse
in the result of non-resonant transitions can be obtained by summation
of $P_{mm'}$ over $m'$. Since each state $|m'\rangle$ differs from
the state $|m\rangle$ by the flip of one $k'$ spin, one can
replace the summation over the states $m'$ to the summation over
the spins $k'\ne k$. For example, the total probability,
$\mu_{N-1}$ (here the subscript of $\mu$ stands for the number of the
resonant spin, $k=N-1$),
of generation of unwanted states by the initial $\pi/2$-pulse is,
\begin{equation}
\label{Pnr1}
\mu_{N-1}=\mu
\sum_{k'=0}^{N-2}\frac 1{|N-1-k'|^2},\qquad\mu=\left(\Omega\over 2\delta\omega\right)^2.
\end{equation}
After the initial $\pi/2$-pulse (which creates
a superposition of two states with equal probabilities (\ref{34})
from the ground state)
the probability of the correct result
is ${\mathcal P}_{\pi/2}=1-\mu_{N-1}$ and the probability of error is
$P_{\pi/2}=1-{\mathcal P}_{\pi/2}=\mu_{N-1}$.
After the first $\pi$-pulse the probability of error becomes,
\begin{equation}
\label{P2}
P_1=1-\frac 12(1-\mu_{L-1})(1-\mu_{L-2}-\varepsilon)-
\frac 12(1-\mu_{L-1})(1-\mu_{L-2}).
\end{equation}
Here we suppose that the values of $\varepsilon=\varepsilon_n$ are the
same for all pulses, and $\Omega_n=\Omega$ for $n\ne 3$ and
$\Omega_3=2\Omega$.

The probability of error in implementation of the whole
logic gate by $\pi/2$-pulse and all $2N-3$ $\pi$-pulses is,
$$
P=P_{2N-3}=1-\frac 12(1-\mu_{N-1})(1-\mu_{N-2}-\varepsilon)
(1-4\mu_{N-2}-\varepsilon)(1-\mu_0-\varepsilon)
\prod_{i=1}^{N-3}(1-\mu_i-\varepsilon)^2-
$$
\begin{equation}
\label{Pnr}
\frac 12(1-\mu_{N-2})(1-4\mu_{N-2})\prod_{i=0}^{N-3}(1-\mu_i)^2,
\end{equation}
where $\mu_i$ can be derived from (\ref{Pnr1}) changing $N-1$ to $i$.
Two last terms in (\ref{Pnr}) are connected with two terms in (\ref{36}).
The probability of non-resonant transitions generated by the third pulse
is approximately four times larger (terms with the factor $4$ in
(\ref{Pnr})) than other probabilities because the Rabi
frequency for this pulse is larger, $\Omega_3=2\Omega$ (to keep
$\varepsilon$ the same). Eq. (\ref{Pnr}) takes into account
all near-resonant transitions, characterized by the parameter $\varepsilon$,
and non-resonant transitions, characterized by the parameters $\mu_i$.
There is no need to consider the phases of the unwanted states,
it is enough to consider only their probabilities.

Eq. (\ref{Pnr}) is convenient for choosing optimal parameters
for application to the scalable solid state quantum computer with a large
number of qubits. On the one hand, the number of qubits is
a scalable parameter in (\ref{Pnr}), which can be easily
changed and increased. (For example one can put $N=1000$.) On the other
hand, our approach takes into consideration all significant
near-resonant and non-resonant
processes generating errors in the quantum logic gates. The parameter
$\varepsilon$ can be minimized (down to the value $\varepsilon=0$) by
the $2\pi k$-method for any number of qubits. However, another
parameter, $\mu$, can be significantly decreases only by increasing
the gradient of the magnetic field, or $\delta\omega$. Since the
number of non-resonant transitions is large, their contribution to
the probability of errors quickly increases with $N$ increasing.
As a consequence, for given $\delta\omega$ there is a restriction on the
number of qubits in our computer, $N<N_{max}$, if we want to keep
the errors below some definite threshold (see section \ref{sec:numerical}).

\begin{figure}[t]
\vspace{-9mm}
\centerline{\psfig{file=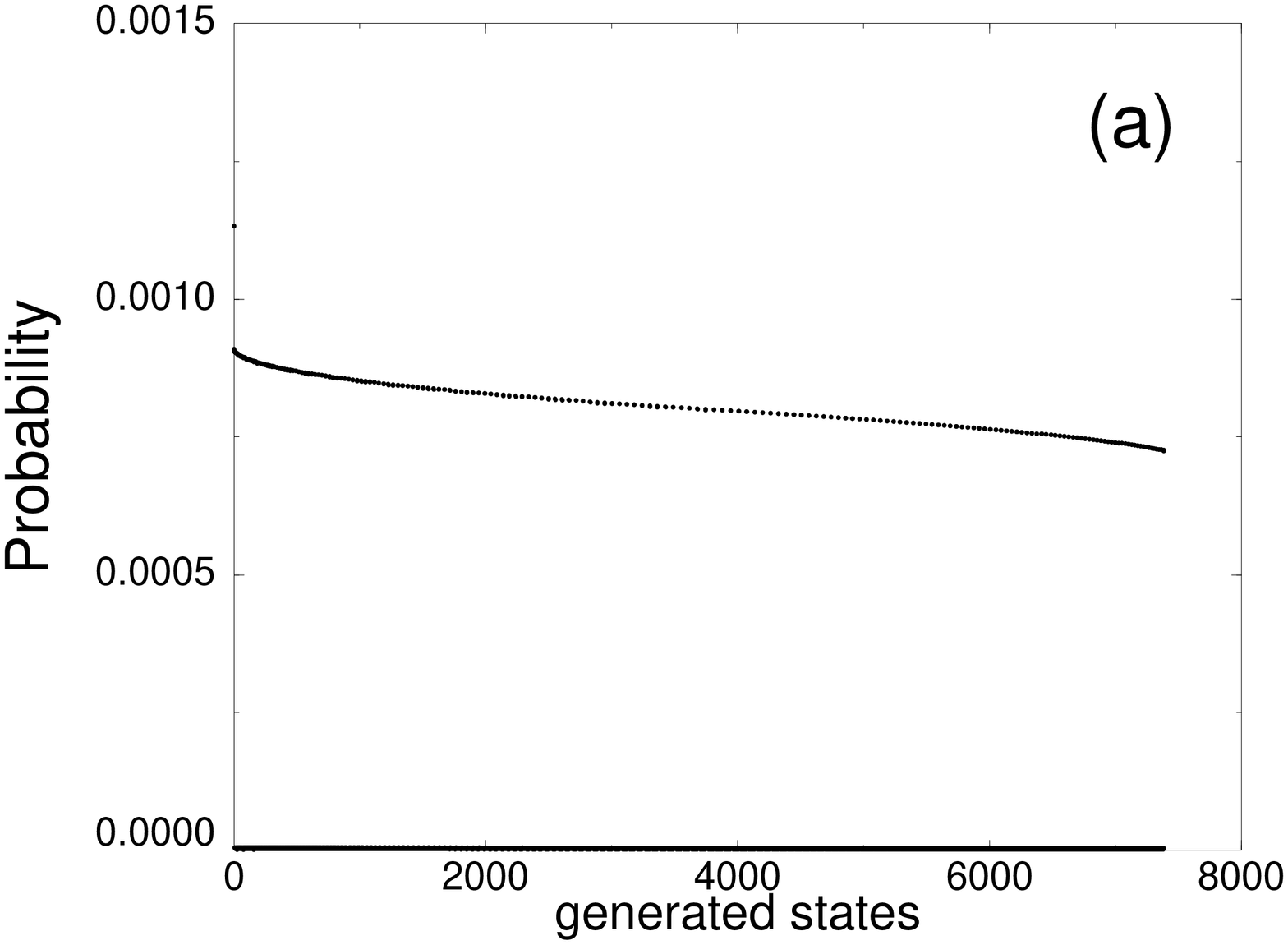,width=80mm,height=61mm,angle=-90}}
\vspace{-6mm}
\centerline{\psfig{file=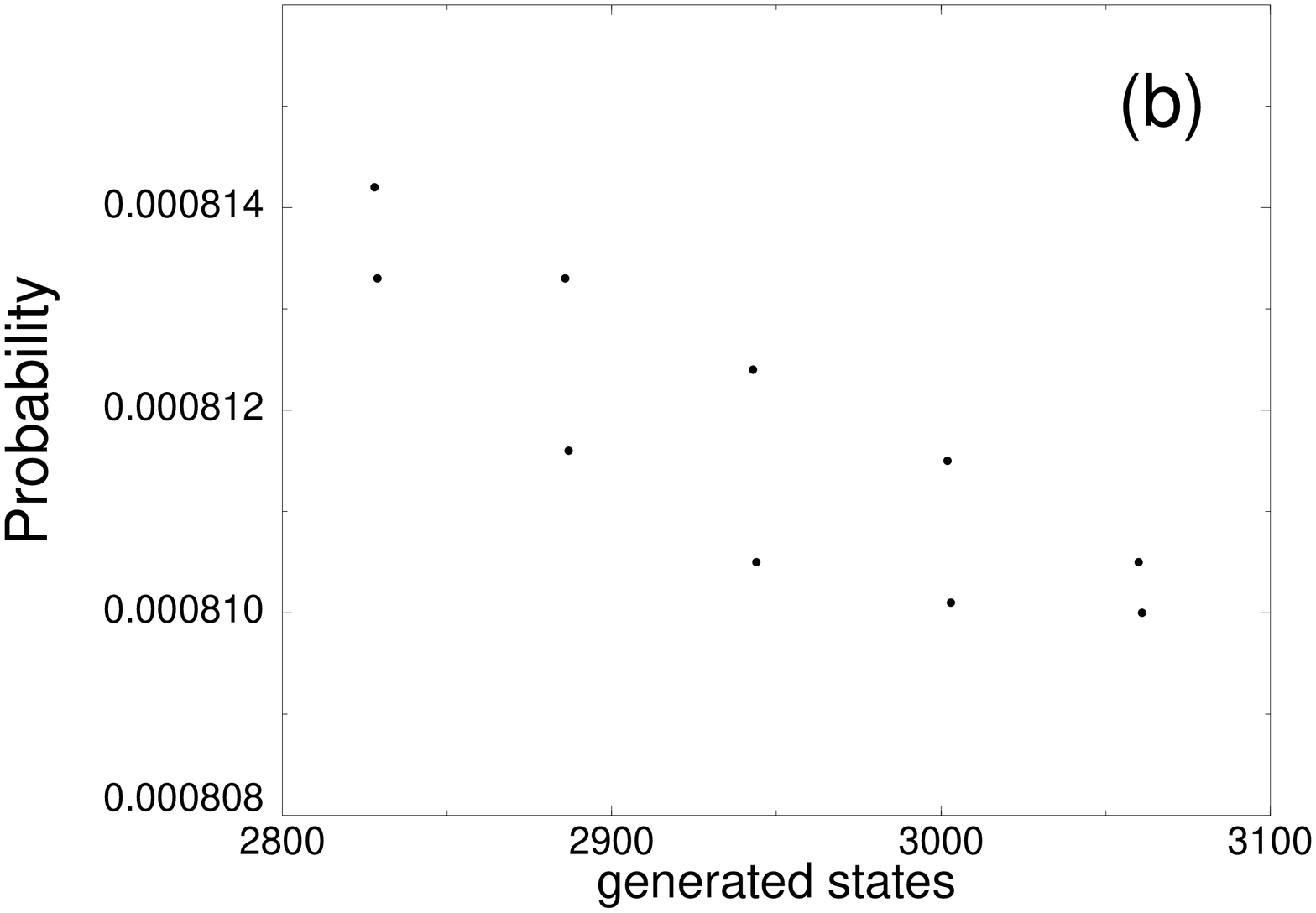,width=80mm,height=61mm,angle=-90}}
\vspace{-6mm}
\centerline{\psfig{file=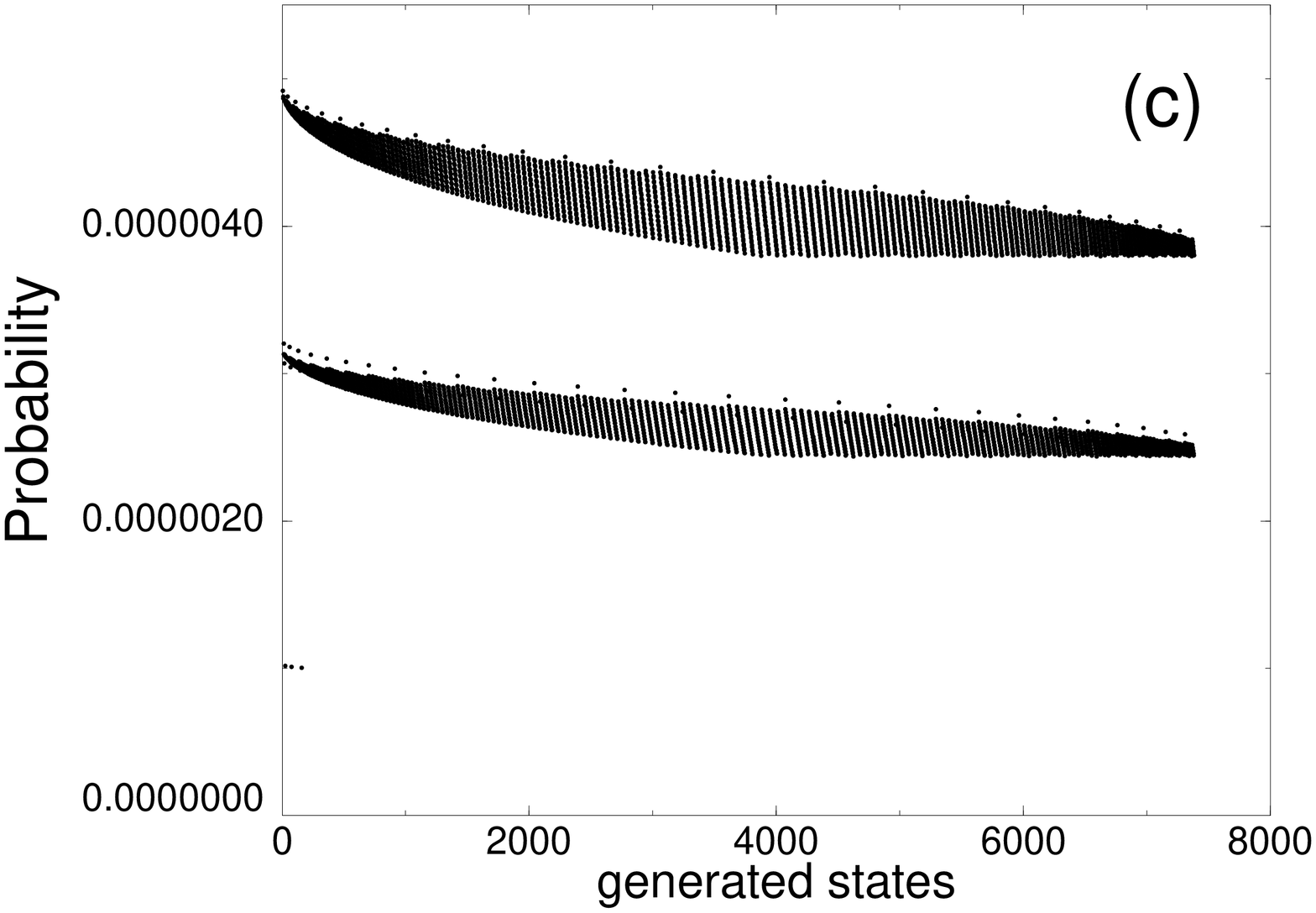,width=80mm,height=61mm,angle=-90}}
\vspace{-2mm}
\caption{Probabilities of unwanted states. The total number of qubits:
$N=200$; $\Omega=0.14$. (The value of $\Omega$ is measured in units of
$J$.) The number of unwanted states with probabilities $|C_n|^2\ge 10^{-6}$
is $7385$. The unwanted states are presented in the order of their
generation. (a)~Two ``bands'': the upper band and the lower band;
(b)~The upper band of Fig. 2a, magnified;
(c)~The lower band of Fig. 2a, magnified.}
\label{fig:2}
\end{figure}

\section{Computer simulations for finite $\varepsilon$}
\label{sec:simulat}
Here we report our results on computer simulations of a quantum
$CN$ gate for remote qubits in quantum computer with 200 and 1000 qubits.
We have developed a numerical code which allows us to study the dynamics
of all quantum states with the probabilities no less than $10^{-6}$ for
a spin chain with up to 1000 qubits.
All frequencies in this and next sections are dimensionless and
measured in units of $J$.
The values of $\Omega$ for the results presented in this section are
the same for all pulses, $\Omega_n=\Omega$, so that
$\varepsilon_n=\varepsilon(\Omega/2J)=\varepsilon$ for $n\ne 3$ and
$\varepsilon_3=\varepsilon(\Omega/4J)$. All probabilities in this section
are doubled. The sum of the probabilities of all these states was
$2-O(10^{-6})$ (the normalization condition). In this section we suppose
that $\varepsilon$ is large, $\varepsilon\gg\mu$
(we suppose $\mu=0$ and $\delta\omega=\infty$),
and one can neglect the non-resonant transitions, since their
contribution to the probability of errors is small. The numerical results
for the case of finite $\mu$ are presented in the next section.

\begin{figure}
\centerline{\psfig{file=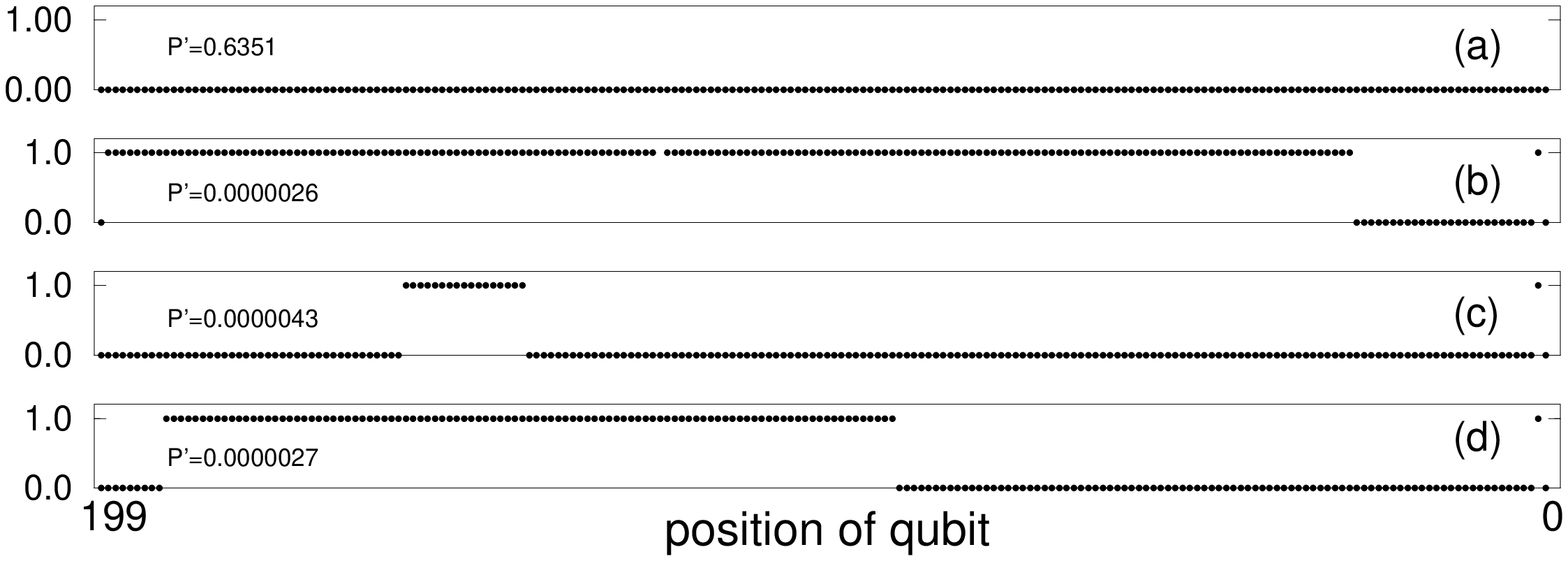,width=95mm,height=65mm,angle=-90}}
\vspace{-3mm}
\caption{(a) The ground state of the spin chain; (b-d) Typical unwanted
states with probabilities $P'\ge 10^{-6}$. The horizontal axis shows the
position of a qubit in the spin chain of $N=200$ spins. The vertical axis
shows the ground state, $|0\rangle$, or the excited state, $|1\rangle$, of
the qubit. Examples of ``high energy'' (b), ``low energy'' (c), and
``intermediate energy'' (d) unwanted states of the whole chain.}
\vspace{3mm}
\centerline{\psfig{file=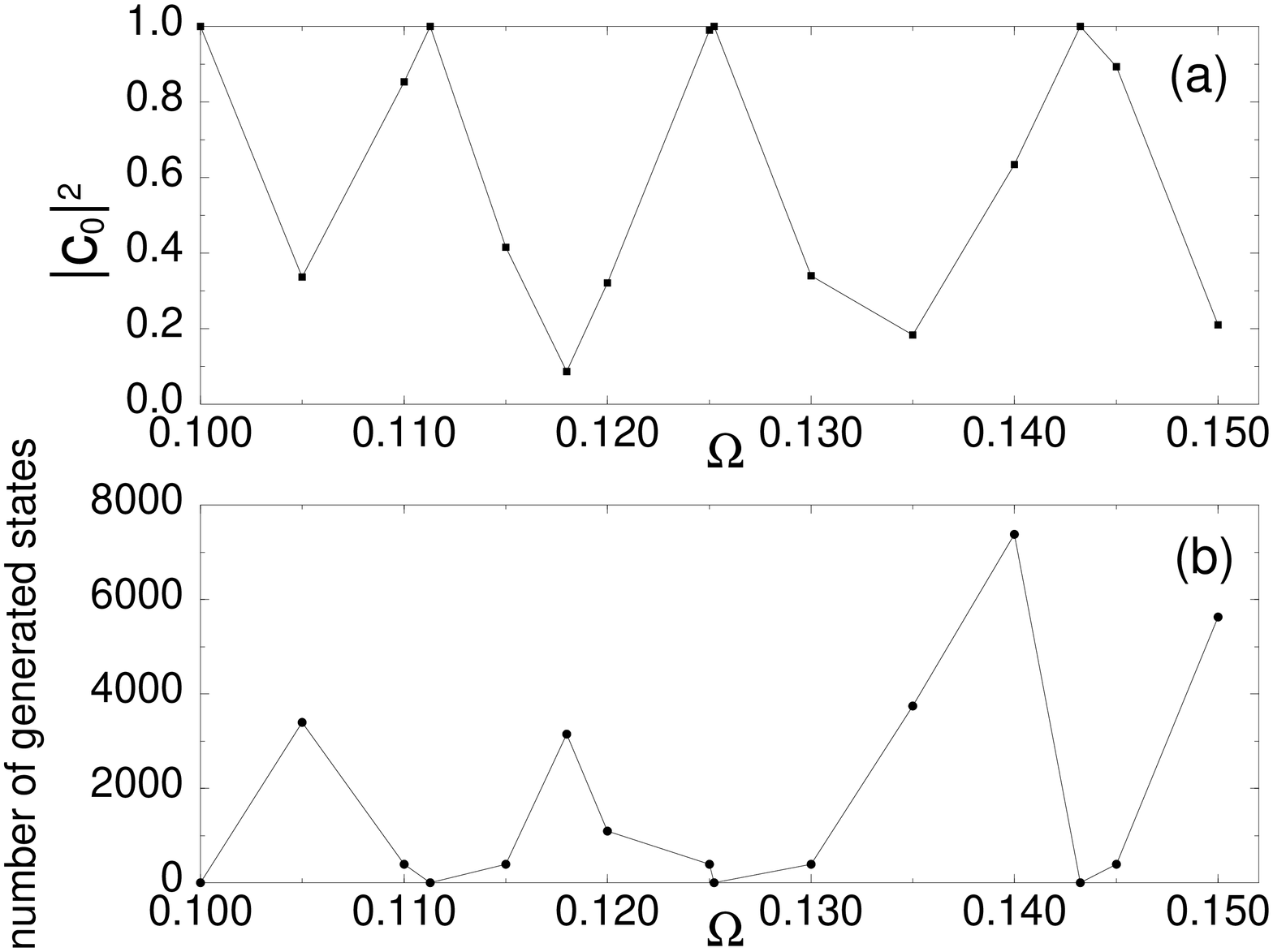,width=85mm,height=65mm,angle=-90}}
\vspace{4mm}
\caption{(a) Probability, $|C_0|^2$, as a function
of $\Omega$. (The value of $\Omega$ is measured in units of $J$.)
The total number of qubits, $N=200$; (b) The total number of unwanted
states with probabilities $|C_m|^2\ge 10^{-6}$.}

\label{fig:3_4}
\end{figure}

In Fig. 2a, we show the probability of the excited unwanted states after
implementation of the $CN_{199,0}$ gate, for $N=200$ and $\Omega=0.14$.
Because we chose $|\Delta|=2J$, it follows from (\ref{28})
that the closest value of the Rabi frequency, $\Omega$, which satisfies
the $2\pi k$ condition is:
$$
\Omega^{[7]}={{2}\over{\sqrt{195}}}\approx 0.1432.
$$
This value differs slightly from the one used in the simulations whose
results are presented in Fig. 2a ($\Omega=0.14$) by $3.2\times 10^{-3}$.
At $\Omega=\Omega^{[7]}$ there are no the near-resonant transitions
and the probability of errors is of order of $\mu$, i.e. very small.
One can see in Fig. 2a that even this small deviation from the $2\pi k$
condition results in generating over 7000 unwanted states during the total
protocol. On the horizontal axis in Fig. 2a, the unwanted states are shown
in the order of their generation. A total of $7385$ unwanted states were
generated whose  probability $P'\ge 10^{-6}$. (In all figures 2-4 only
the states with $P'\ge 10^{-6}$ are taken into account.) The probability
distribution of unwanted states clearly contains two ``bands''. One group
of these states has the probability, $P'\sim 10^{-6}$ (the bold ``line''
near the horizontal axis).  The second group of states has the probability
$P'\sim 10^{-3}$ (the upper ``curve'' in Fig. 2a). Fig. 2b shows an
enlargement of the upper ``band'' of the Fig. 2a. One can see some
sub-structure in this ``band''. Fig. 2c shows the sub-structure in
the lower ``band'' of Fig. 2a. Our
simulations show the existence of sub-structure in the upper and lower
``bands'' shown in Fig. 2c. There exists a hierarchy in the structure of
the distribution function of unwanted generated states. Figs. 3(b-d) show
the typical structure of unwanted states of the spin chain. In Figs. 3(a-d)
``0'' corresponds to the ground
state, $|0\rangle$, of a qubit, and ``1'' corresponds to the excited
state, $|1\rangle$, of a qubit. Fig. 3a shows the ground state of the spin
chain. (All qubits are in their ground state, $|0\rangle$.)
$P'$ in Fig. 3,
is the probability of a state of the whole chain.
All unwanted states in
Figs 3(b-d) belong to the upper and lower ``bands'' shown in Fig. 2c, and
they have probabilities, $P'\sim 10^{-6}$.


\begin{figure}
\centerline{\psfig{file=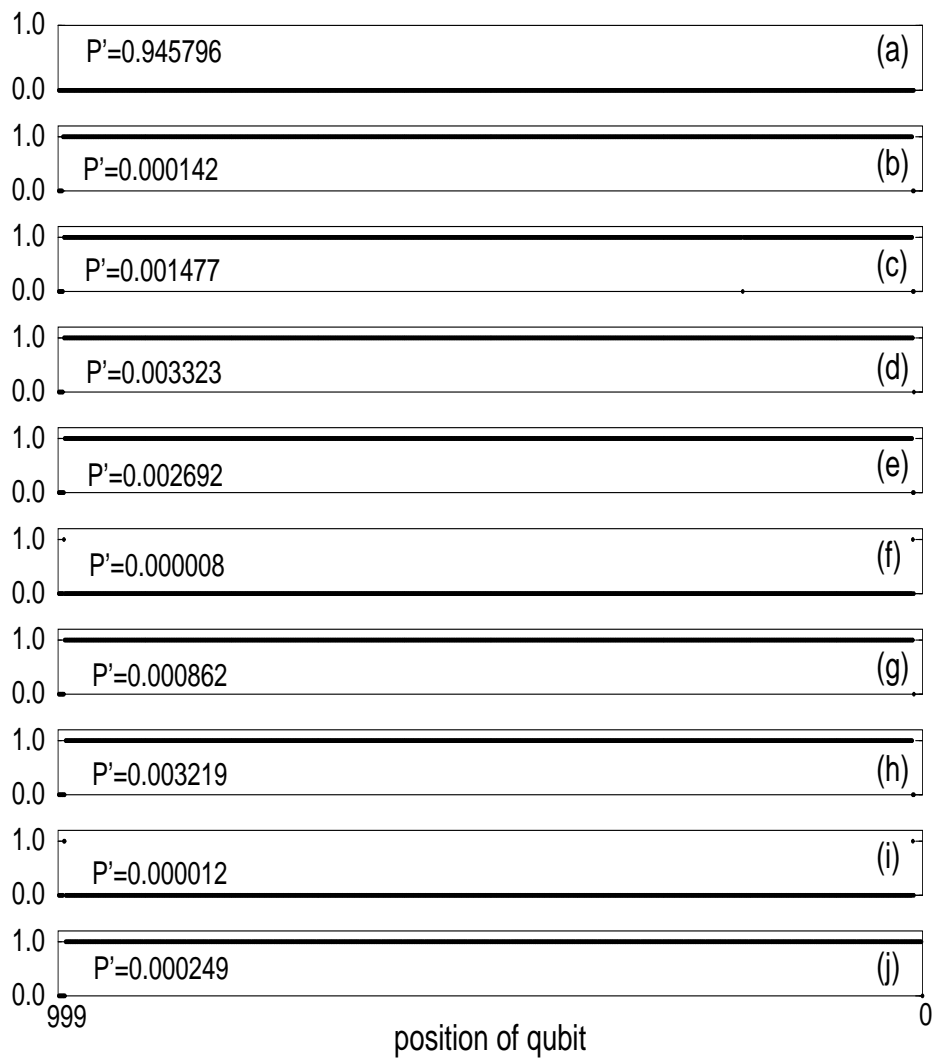,width=95mm,height=125mm,angle=-90}}
\vspace{10mm}
\caption{(a) The ground state of the chain; (b-j) Examples of unwanted
states ($N=1000$); $\Omega=0.1$ for all $\pi$-pulses except for the
$\pi$-pulses from the 10-th to the 40-th for which $\Omega=0.1+\eta$,
where $\eta$ is a random parameter in the range: $-0.05<\eta<0.05$.}
\label{fig:5}
\end{figure}

\noindent
Note, that many unwanted states
are the high energy states of the spin chain (many-spin excitations).
Typical unwanted states contain highly correlated spin excitations.
Fig. 4 shows the probability of the ground state, $|C_0|^2$, and the
total number of unwanted states (with probability $P'\ge 10^{-6}$) as
a function of the Rabi frequency, $\Omega$.
(The point $\Omega=0.14$
corresponds to the results shown in Fig. 2.) The maximum value of
$|C_0|^2$ and the minimal total number of unwanted states correspond to
values of $\Omega$ which satisfy the $2\pi k$ condition for $396$ of the
total number of $\pi$-pulses, $397$.
(The third $\pi$-pulse does not satisfy the $2\pi k$ condition.)
One can see from Fig. 4 that application of the $2\pi k$ condition
significantly
improves the performance of the quantum Control-Not gate. Fig. 5 shows
the ground state and the examples of unwanted states for $N=1000$.
In this case, for all $\pi$-pulses $\Omega=0.1$, except for the
$\pi$-pulses from 10-th to 40-th for which $\Omega=0.1+\eta$, where
$\eta$ is a random number in the range: $-0.05<\eta<0.05$. The value
$\Omega=0.1$ corresponds to the $2\pi k$ condition (\ref{35}) with $k=10$.
One can see from Fig. 5 that the states of the quantum computer are
strongly correlated and some of them are highly excited. To avoid these
unwanted strongly correlated and highly excited states it will be necessary
to apply both a $2\pi k$-method and adequate error correction codes.

\section{Numerical results for total error}
\label{sec:numerical}
In the previous section we presented the numerical results when the
conditions of the $2\pi k$-method are not satisfied and
$\varepsilon\gg\mu$. In this case one can neglect the non-resonant
transitions and calculate the error caused by the near-resonant transitions
by Eq. (\ref{41}) or (\ref{41a}).
One can minimize these errors by choosing $\varepsilon_n=0$ for all $n$
($2\pi k$ condition).
We show in this section that even under the conditions of the
$2\pi k$-method the error can be relatively large due to
non-resonant transitions, and the probability of errors increases
when the number of qubits, $N$, increases.

We present here the results of computer simulations
taking into consideration both, near-resonant and non-resonant transitions.
We test our approximate formula (\ref{Pnr}) by exact numerical solution
using Eqs. (\ref{dynamics}) and (\ref{AC}) when the number of qubits is
not very large ($N=10$), so that the total number of states in the Hilbert
space is $2^{10}=1024$. As before, the frequencies are dimensionless and
measured in units of $J$.
The values of $\varepsilon$ for the results presented in this section are
the same for all pulses, $\varepsilon_n=\varepsilon$, so that
$\Omega_n=\Omega$ for $n\ne 3$ and
$\Omega_3=2\Omega$. The norm of the wave function is equal (as usual)
to unity and $\delta\omega$ (and $\mu$) is finite.

In Figs. 6(a,b) we compare the
total probability, $P$,
of generation of unwanted states, found from the analytical
estimate (\ref{Pnr}), with the result of exact numerical
solution for the chain containing $N=10$ spins.
As follows from these figures, there is a very good
correspondence between our approximate and exact solutions.
From Fig.~6a one can see that at $\varepsilon=0$ the value of
$P$ has tendency to decrease as $const/\delta\omega^2$.
In Fig. 6b we plot the probability,
$P$, as a function of $\Omega$.
When $\varepsilon\gg\mu$,
and $J\ll\delta\omega$, the probability of generation of unwanted
states is mostly defined by the value of $\varepsilon$
and is almost independent of $\delta\omega$ since $\varepsilon\gg\mu$.
The value
of $\delta\omega$ fixes the values of the minima
in Fig. 6b: for larger
$\delta\omega$ the minima in the plot in Fig. 6b become deeper.
The values of different minima in Fig. 6b indicate the contribution
of non-resonant processes to the
probability $P$. Since the value of $\Omega$ in Fig. 6b
does not change significantly, the contribution of non-resonant
processes to the probability of errors is approximately the same for all
$\Omega$.
One can see that this contribution is negligible in comparison
with the contribution of the near-resonant processes
(defined by $\varepsilon$)
for all $\Omega$, except for the small regions of $\Omega$, where
$\varepsilon=0$.

\begin{figure}
\centerline{\mbox{\psfig{file=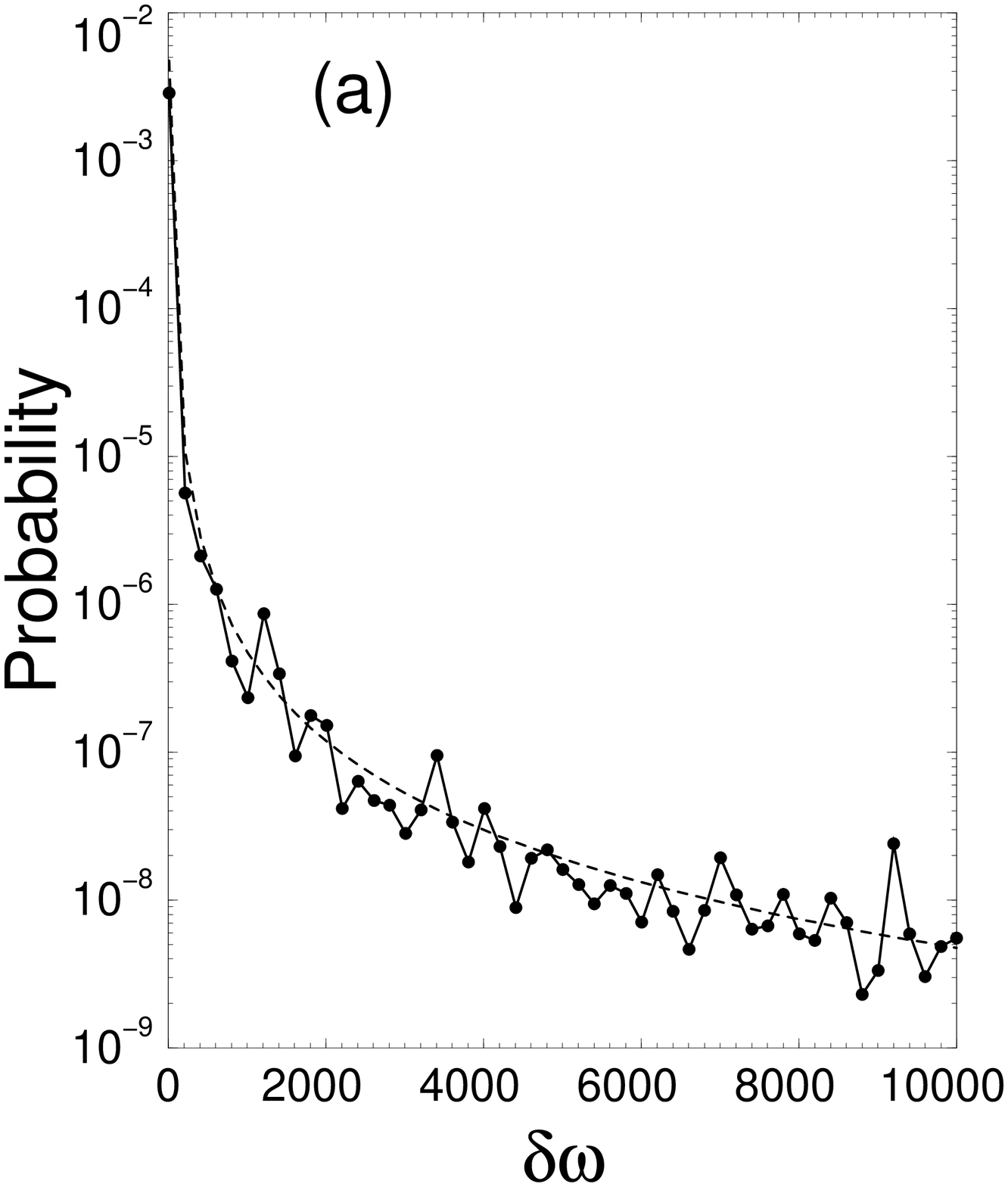,width=6.2cm,height=7cm}\hspace{-3mm}
       \psfig{file=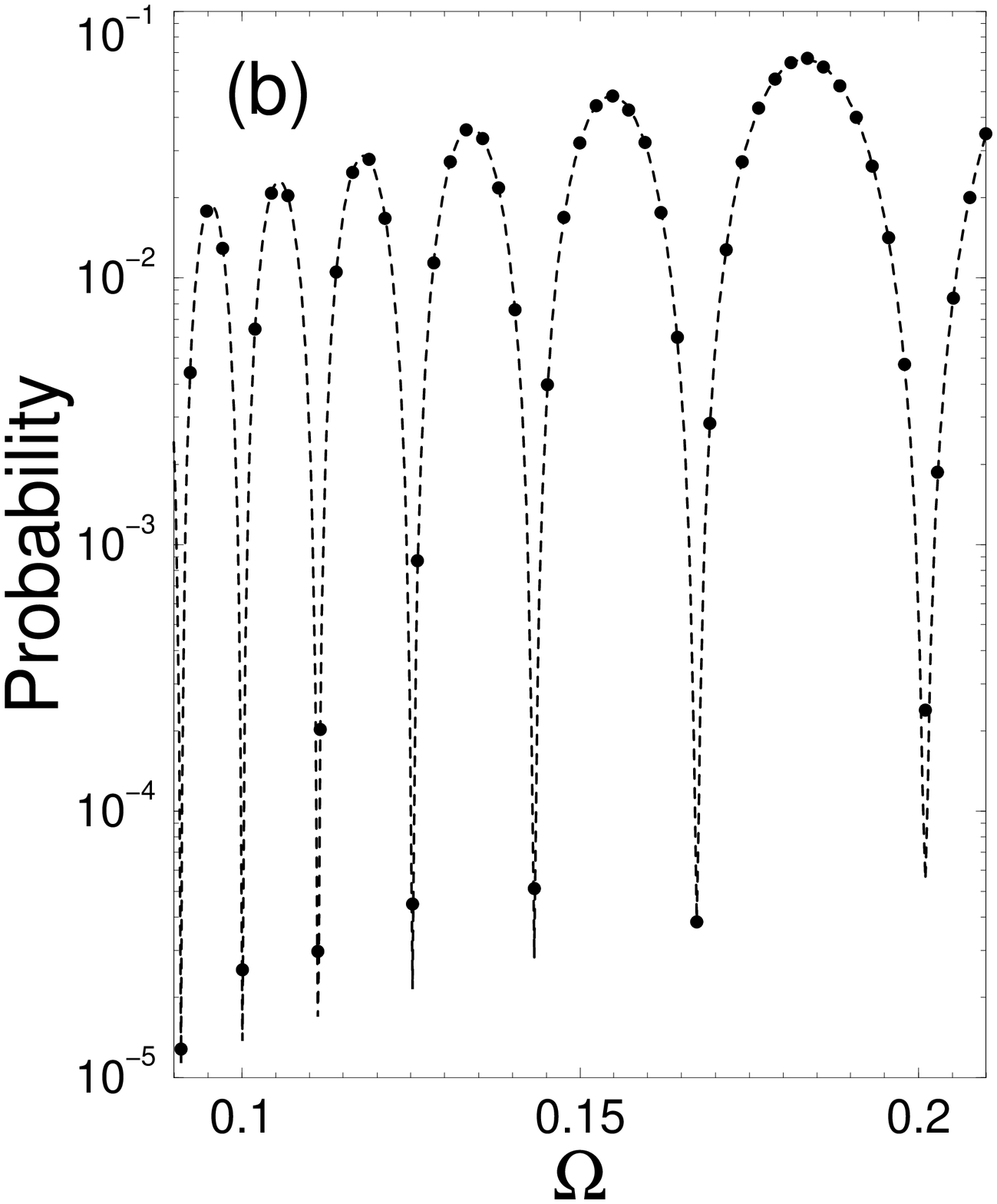,width=6.2cm,height=7cm}}}
\vspace{-3mm}
\caption{The probability, $P$,
of generation of unwanted states. Filled
circles are the results of exact numerical solution,
dashed lines are the analytic estimates (\ref{Pnr}).
(a) $P$ as a function of $\delta\omega$,
$\varepsilon=0$, $\Omega=\Omega^{[8]}$ from (\ref{Omega_k}).
(b) $P$ as a function of $\Omega$, $\delta\omega=100$. $N=10$}

\vspace{2mm}
\centerline{\mbox{\psfig{file=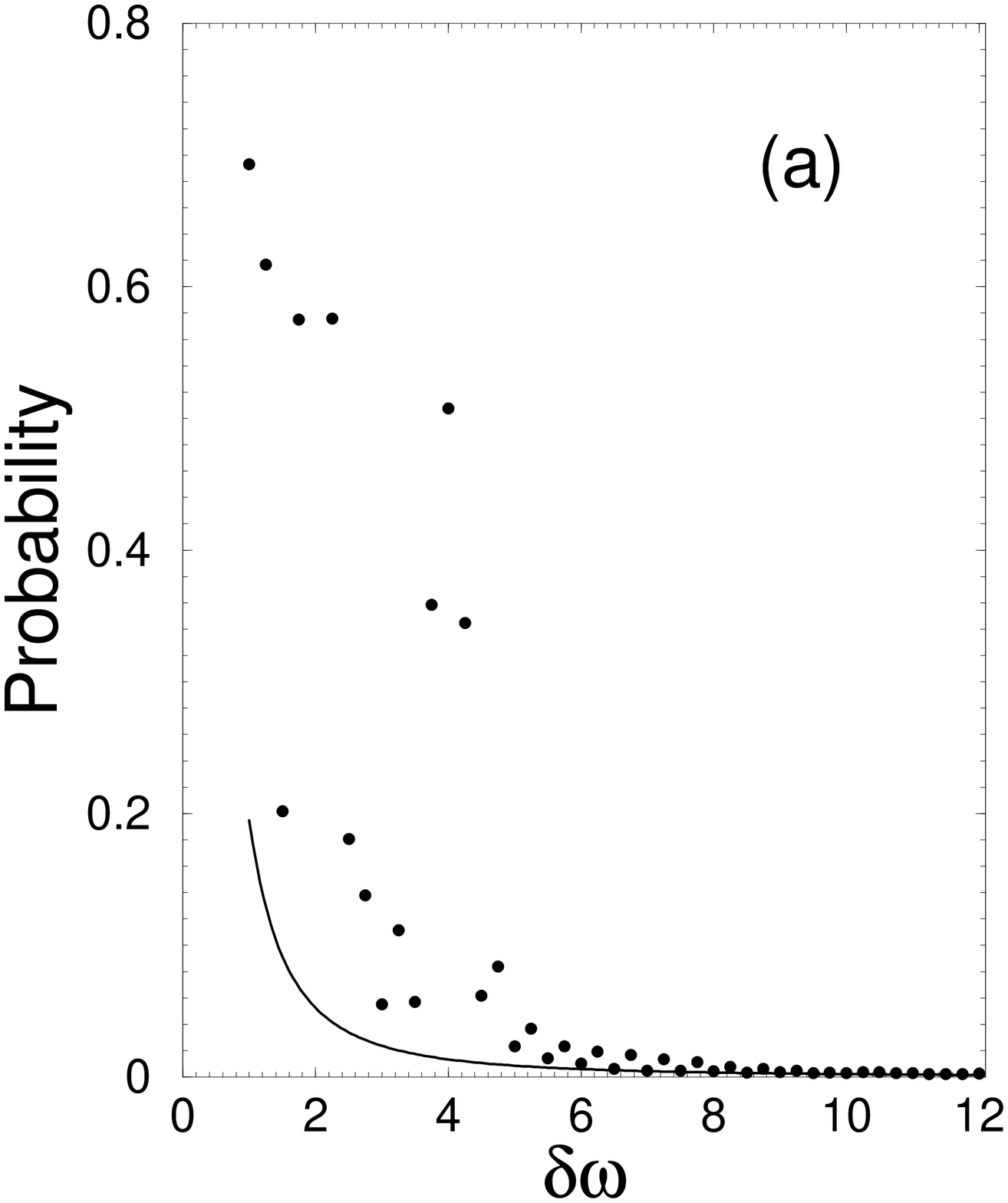,width=6.0cm,height=7cm}
       \psfig{file=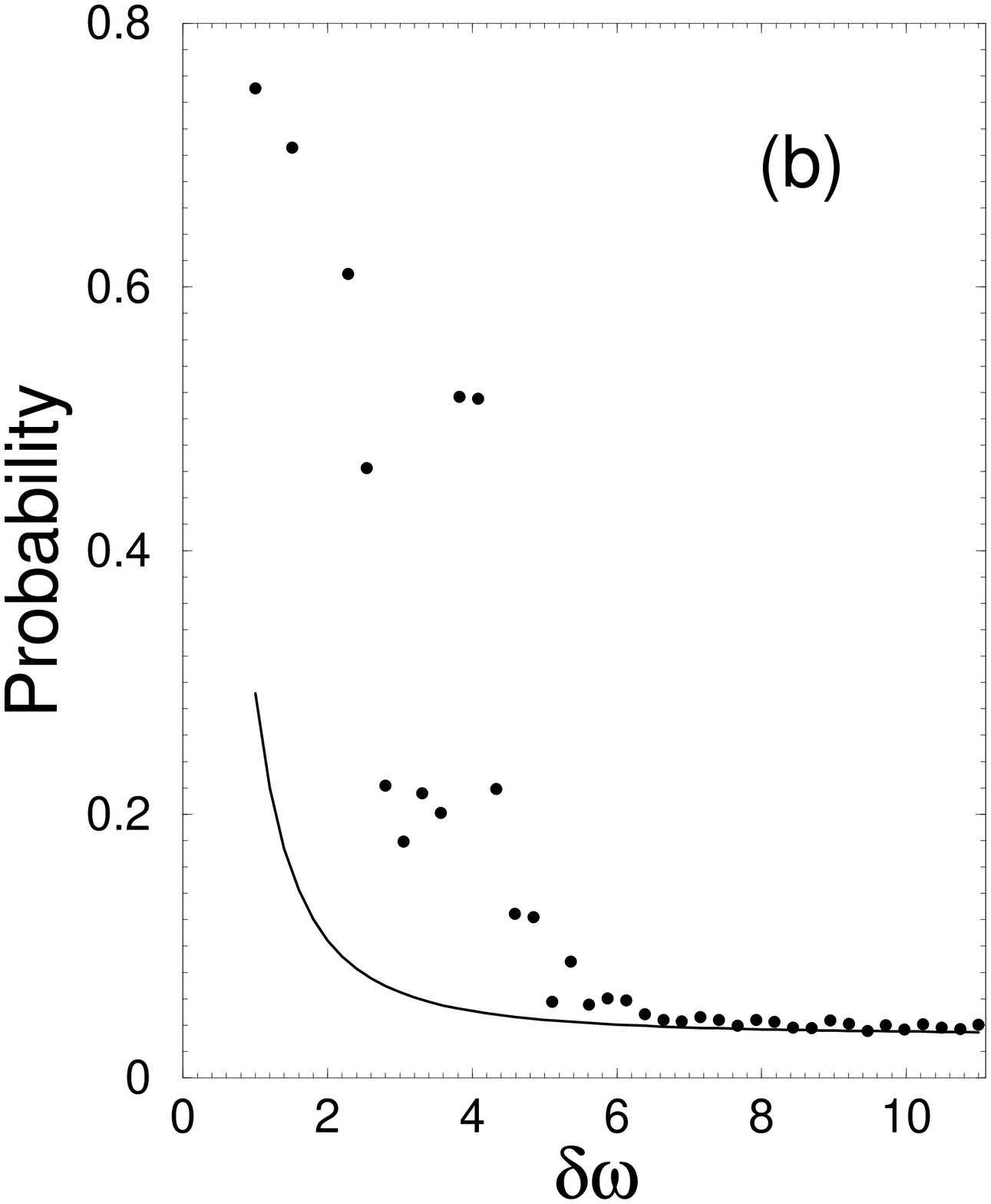,width=6.0cm,height=7cm}}}
\vspace{-3mm}
\caption{The total probability, $P$, of unwanted states
as a function of $\delta\omega$ when the value of
$\delta\omega$ is comparable with the value of the interaction constant,
$J$.
(a) $\Omega=\Omega^{[8]}$ ($\varepsilon=0$), (b) $\Omega=0.15$
($\varepsilon=0.0039$).
Filled circles are the results of exact numerical solution,
solid line is the analytical estimate (\ref{Pnr}). $N=10$.}
\label{fig:6_7}
\end{figure}

We should note that one more condition (except for $\varepsilon,\,\mu\ll 1$)
must be satisfied for Eq. (\ref{Pnr}) to be valid. The value
of the interaction constant, $J$, should be small in comparison with the
difference between the spin frequencies, $J\ll\delta\omega$.
In Figs. 7a and 7b we plot
the probability, $P$, as a function of
$\delta\omega$ for $\varepsilon=0$ and $\varepsilon\not=0$.
(Since the frequencies are measured in units of $J$ the value
$\delta\omega=1$ in Figs. 7(a,b) corresponds to $\delta\omega=J$.)
One can see that our results are valid
only when $J\ll\delta\omega$ ($\delta\omega\gg 1$ in Figs. 7(a,b)),
in spite of the fact that the
parameter $J/\delta\omega$ does not appear explicitly
in (\ref{Pnr}). From Fig. 7b one can see that the probability
of unwanted states, $P$, for $\varepsilon\gg\mu$ (for large
$\delta\omega$) becomes
relatively independent of $\delta\omega$. In this case the value
of $P$
is defined by the parameter $\varepsilon$ which does not
depend on $\delta\omega$.

Equation (\ref{Pnr}) is convenient for choosing
the real experimental parameters required for operation of the
solid state quantum computer.
Suppose that we are able to correct the errors with the
probability less than $P_0=10^{-5}$.
Our perturbation theory allows us to calculate the region of parameters
for which the probability of error, $P$,
will be less than $P_0$. In Figs. 8(a-d) we plot the diagrams
obtained using our perturbative approach for $N=10$.
Inside the hatched areas the
probability of generation of unwanted states is less than $P_0$.
We should note that building the plots like those in Figs. 8(a-d)
using exact solution of the problem requires a significant number of
computer time even for relatively small number of qubits ($N=10$).
On the other hand, the number $N$ in our perturbation theory is the
parameter which can be increased without any problems.

\begin{figure}[t]
\mbox{\hspace{-8mm}\psfig{file=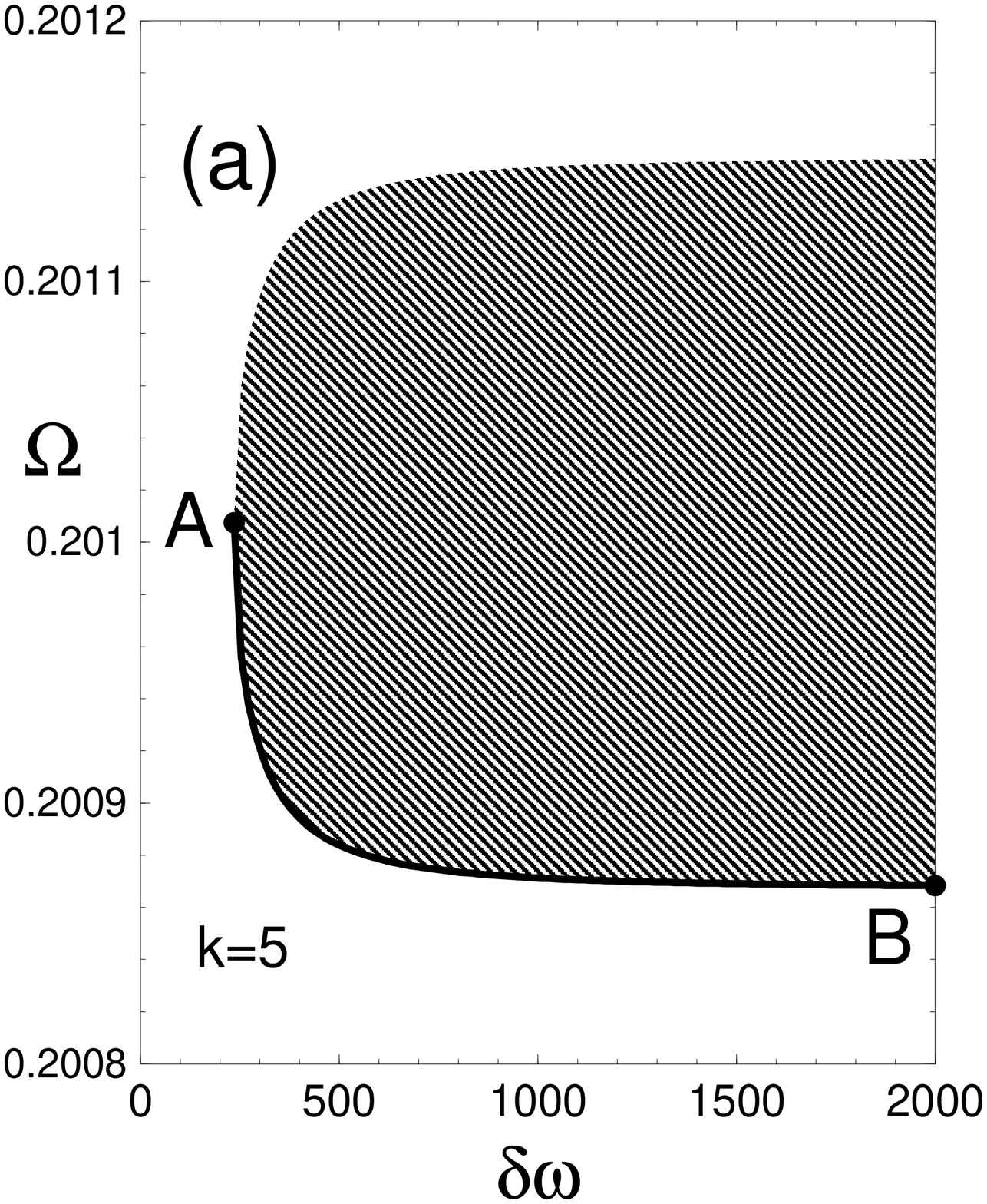,width=6.3cm,height=7.2cm}
\hspace{-4mm}\psfig{file=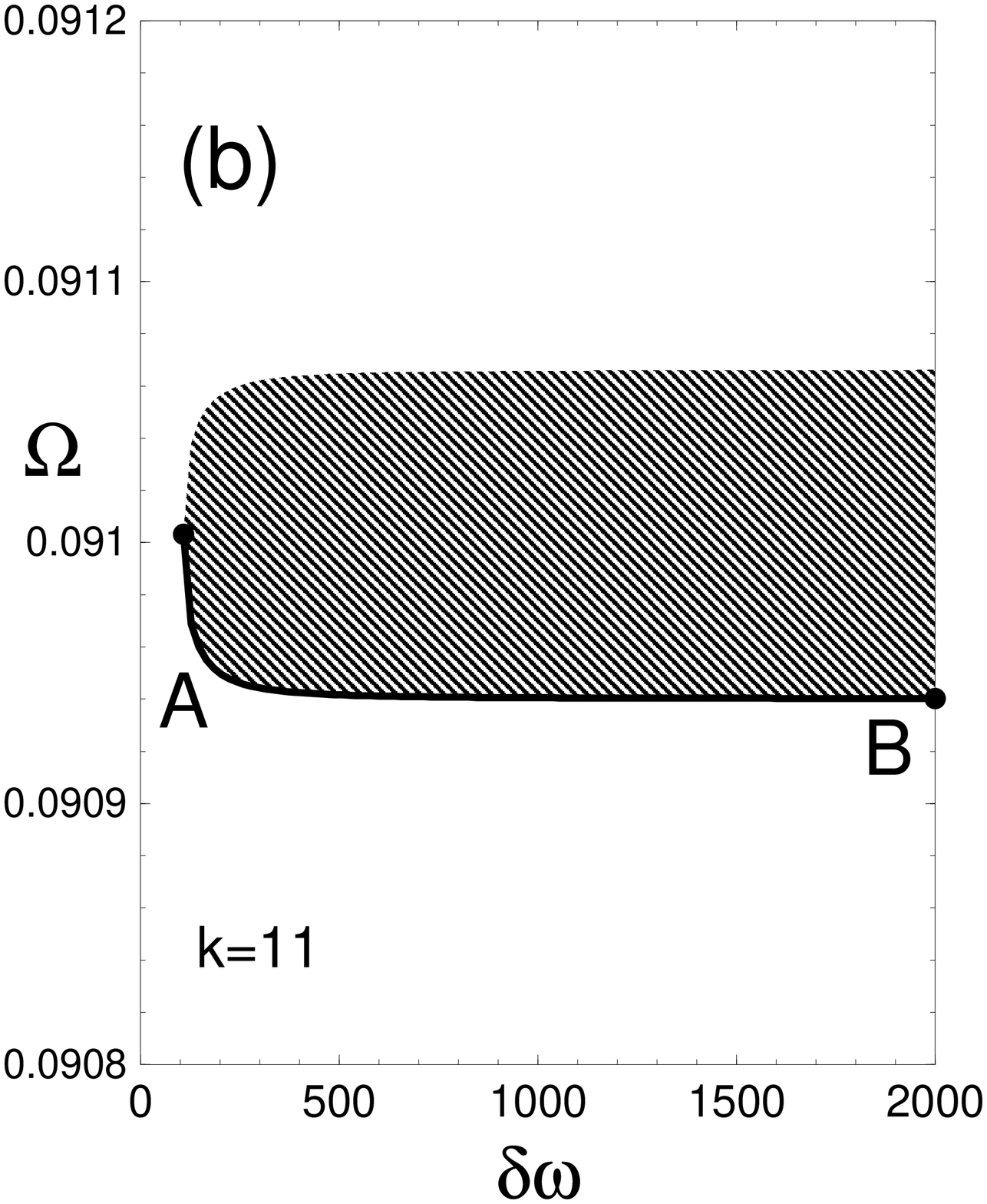,width=6.3cm,height=7.2cm}}
\mbox{\psfig{file=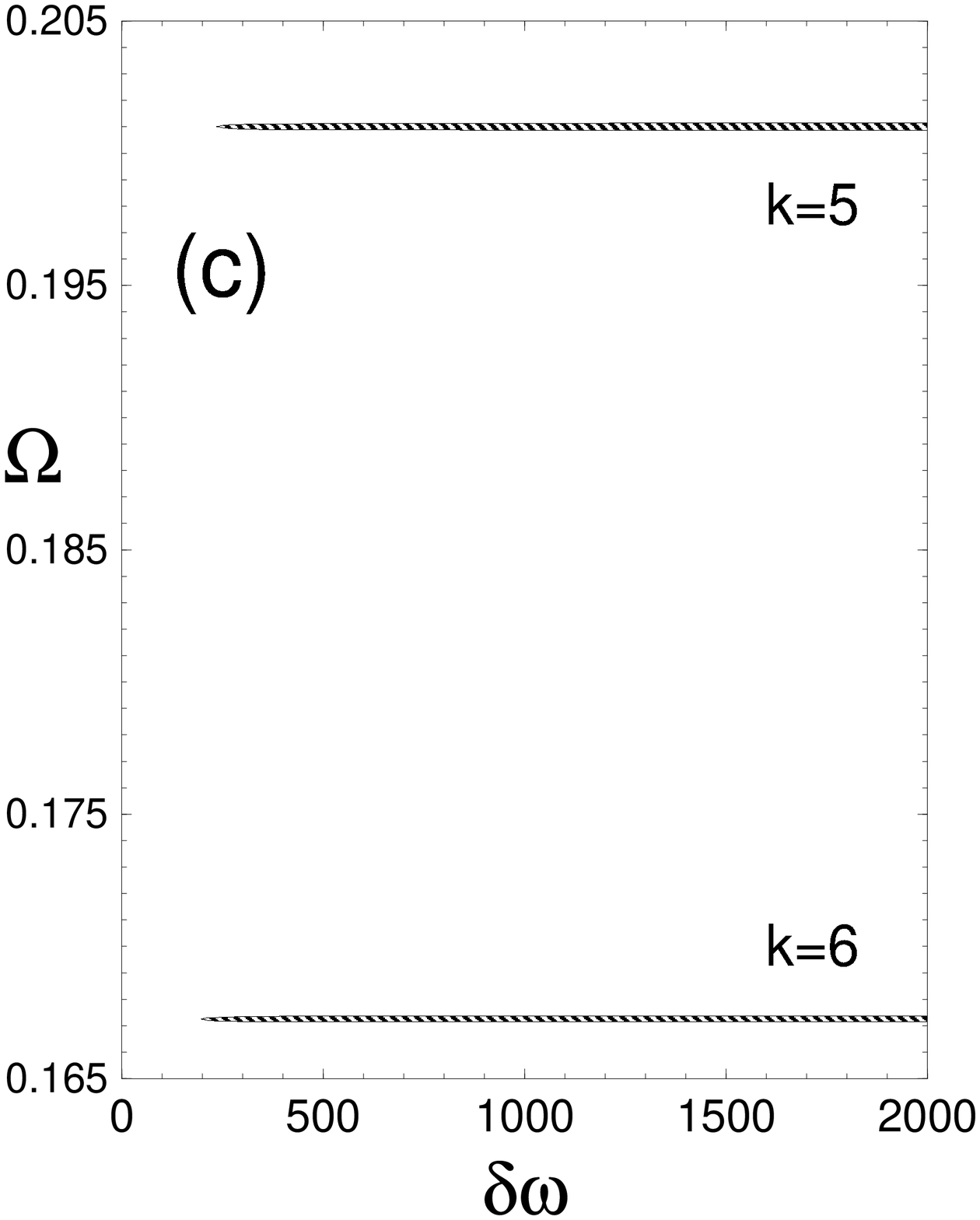,width=6.3cm,height=7.2cm}
\hspace{-4mm}\psfig{file=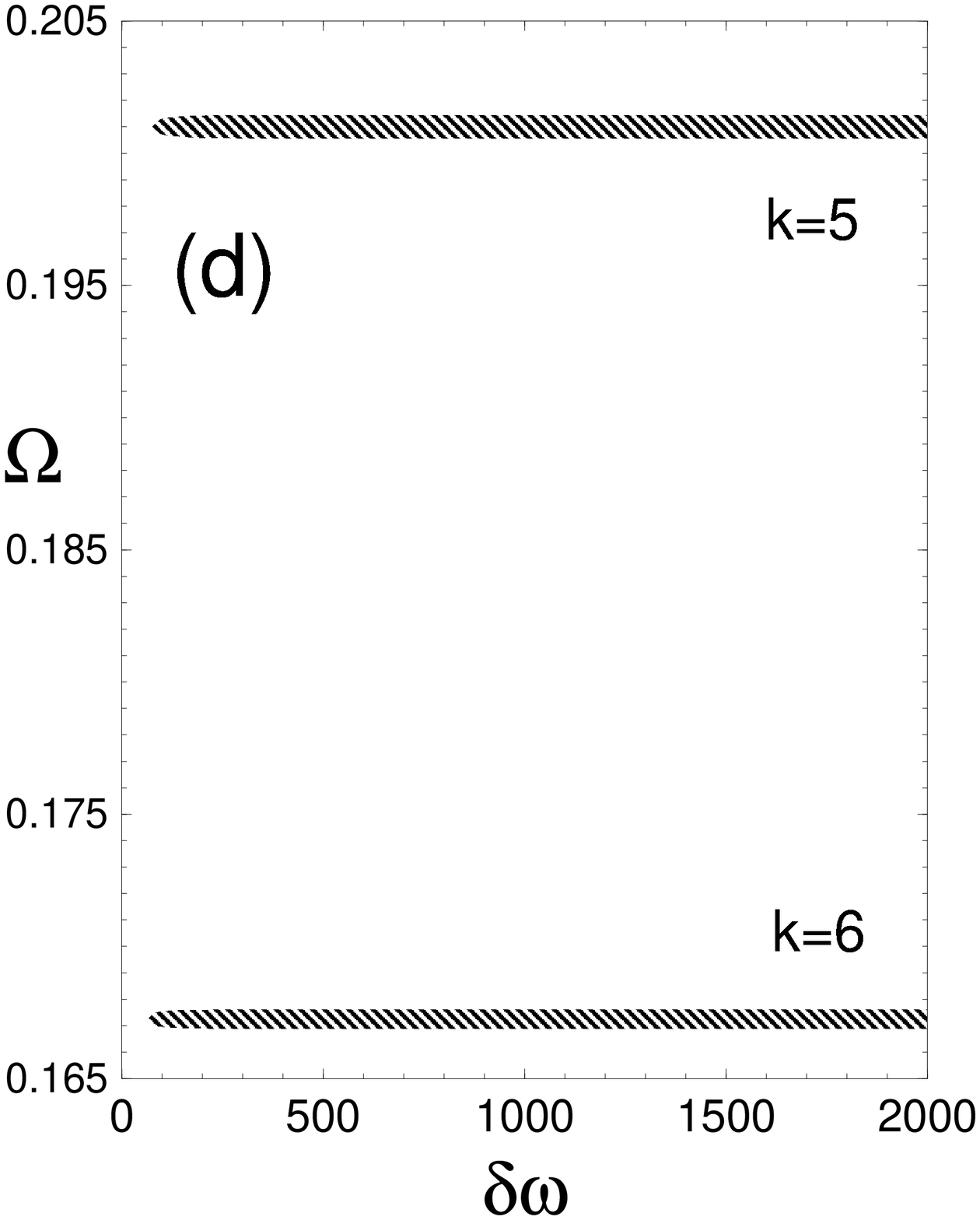,width=6.3cm,height=7.2cm}}
\vspace{-3mm}
\caption{The probability of generation of unwanted states, $P$,
at different values of $\delta\omega$ and $\Omega$. In the hatched
regions $P<P_0$. The different pictures (a) - (d)
indicate the different regions in $\Omega$ in different scales.
(a), (b), (c) $P_0=10^{-5}$; (d) $P_0=10^{-4}$. $N=10$.}
\label{fig:8}
\end{figure}

\begin{figure}[t]
\centerline{\mbox{
\psfig{file=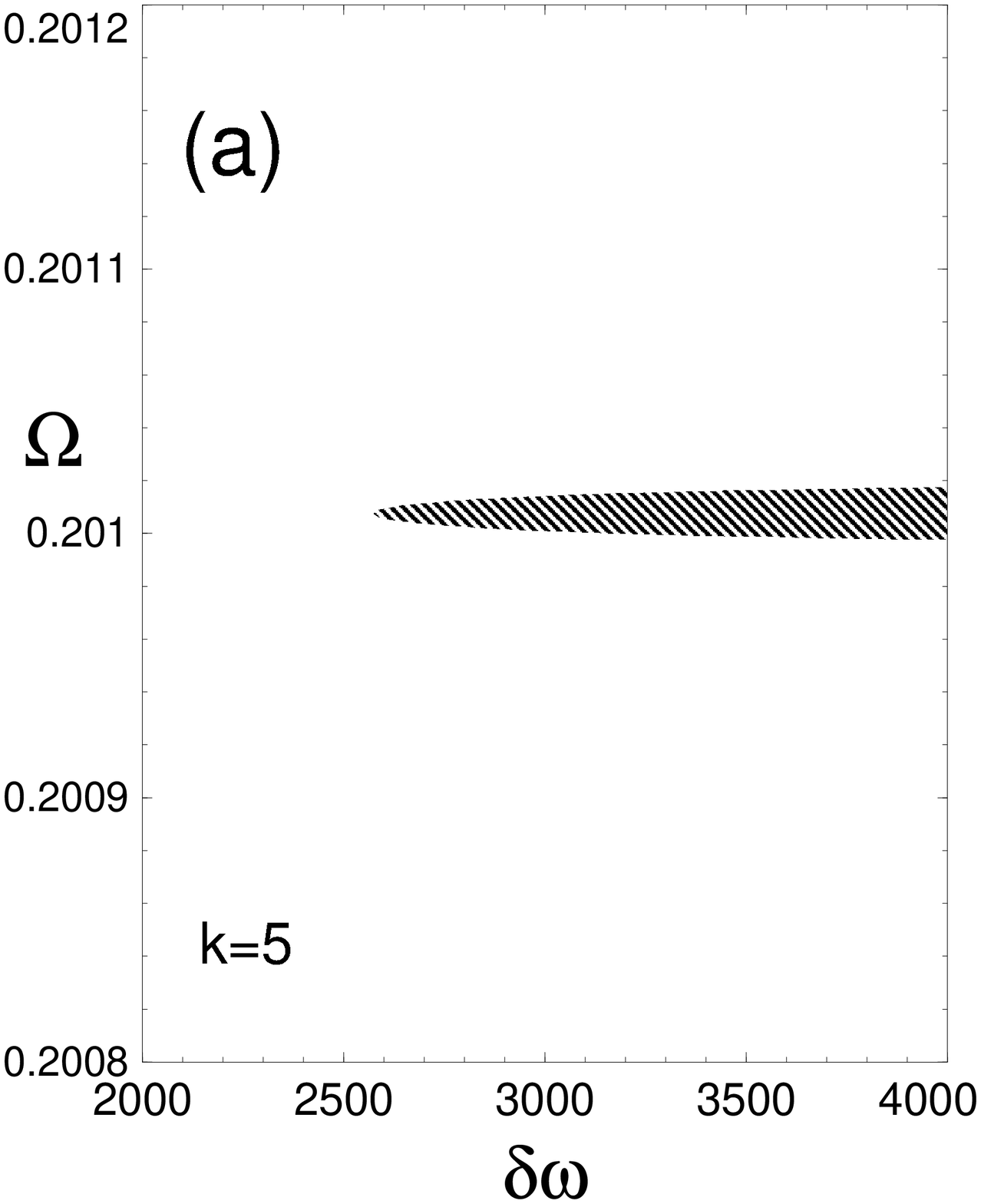,width=6.2cm,height=7.2cm}\hspace{-4mm}
\psfig{file=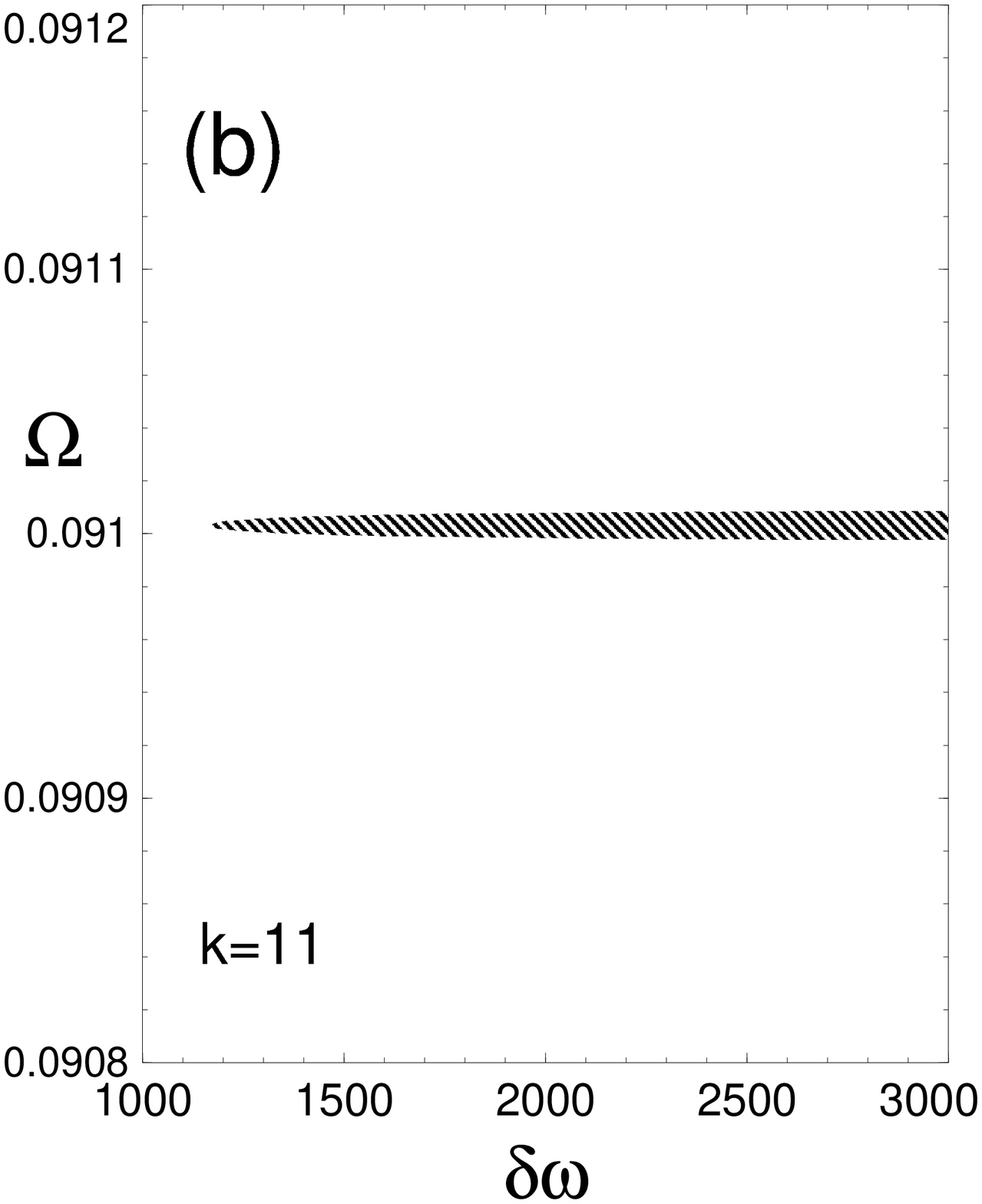,width=6.2cm,height=7.2cm}}}
\vspace{-3mm}
\caption{(a) The same as in Fig. 8a but for $N=1000$.
(b) The same as in Fig. 8b but for $N=1000$. The
scale in $\delta\omega$ shifted to the larger values
in comparison with Figs. 8a, 8b.}
\vspace{3mm}
\centerline{\psfig{file=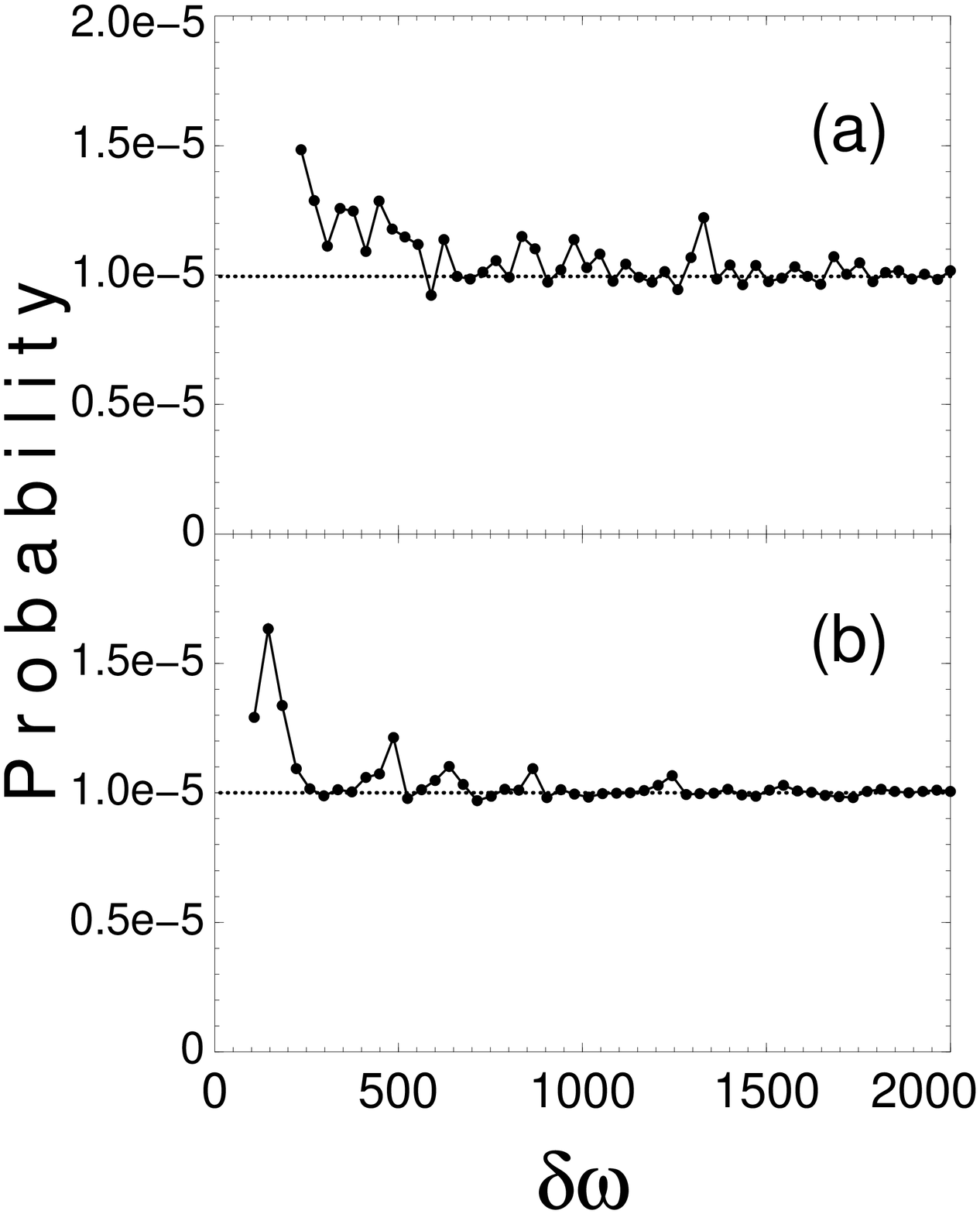,width=11cm,height=8cm}}
\vspace{-4mm}
\caption{The exact solution for the
probability of generation of unwanted states, $P$,
for the parameters, which correspond to: (a) curve AB in Fig.~8a;
(b) curve AB in Fig.~8b. The dashed lines indicate the solution
obtained by the perturbation theory.
$N=10$.}
\label{fig:9_10}
\end{figure}

From Figs. 8(c,d) one can see that
the sizes of the hatched areas are small,
much smaller than
the areas between the neighboring hatched regions.
The total size of the hatched regions is mostly defined by the
probability, $P_0$, of tolerant error which can be corrected
using additional error correction codes.
From comparison of Fig. 8c for
$P_0=10^{-5}$ and Fig. 8d for $P_0=10^{-4}$ one can see that the sizes
of the hatched regions increase with $P_0$ increasing. The
coordinate of the point A in $\Omega$ is $\Omega=\Omega^{[k]}$,
where the values of $k$ are indicated in the figures.

In almost all quantum algorithms the phase of the wave function
is important. We numerically compared the phase of the wave function on the
boundaries of the hatched
regions in Figs. 8(a,b) with the phase in the centers of these regions,
(where $\Omega$ satisfies $2\pi k$ condition,
see Eq. (\ref{37}), $k=5$ and
$k=11$) at fixed $\delta\omega$. The deviation in phase is
only $\sim 0.15\%$.
This is much less that the corresponding change in the probabilities
of errors (by several orders).

In Figs. 9(a,b) we plot the same as in Figs. 8(a,b) but for $N=1000$.
One can see that increasing the number of qubits increases the errors
and the sizes of the hatched areas decrease considerably.
From Figs. 8(a-d) and 9(a-b)
one can see that even when the values of $\Omega$ satisfy
$2\pi k$-method (position of the point A in $\Omega$ in Figs. 8(a,b) and
extreme left points of the hatched regions in Figs. 8(c,d) and 9(a,b))
the error can be relatively large because of the non-resonant processes.
From comparison of Figs. 8(a,b) for $N=10$ with Figs. 9(a,b)
for $N=1000$ one can see that the
minimal value of $\delta\omega$ required to make the errors small
increases considerably with $N$ increasing.

We now analyze the probability of errors as a function of $\delta\omega$.
When the value of
$\delta\omega$ is large enough, the probability of error
(and the widths of the hatched areas in $\Omega$)
becomes practically independent of $\delta\omega$.
This is because at $\delta\omega\gg 1$ and at $\varepsilon\gg\mu$
the error is mostly
defined by $\varepsilon$, which is independent of $\delta\omega$.
As a consequence, one can, for example, estimate
the widths of the hatched areas at $\delta\omega\gg 1$
taking into account only the near-resonant
transitions (formula (\ref{41a})).


In order to test our approximate result we solved the problem
exactly using the eigenstates of the matrices ${\mathcal H}^{(n)}$,
$n=1,\dots M$,
for the parameters which correspond to the lower boundary of the
hatched regions in the Figs.~8a~and~8b (curves AB).
In Figs.~10a and 10b we compare the total probability of errors,
$P$, obtained using the exact and approximate
solutions. One can see again, as in Figs. 6a and 6b, that there is a good
correspondence between the exact and approximate solutions.
A similar correspondence can be demonstrated for other
points of the parametric space ($\delta\omega,\,\Omega$)
when the conditions $\Omega\ll J\ll\delta\omega$ are satisfied.

\section{A classical Hamiltonian for quantum computation}
In this section we demonstrate that the process of quantum
computation, including creation of the entanglement, the dynamics of
quantum controlled operations, and the dynamics of complicated quantum
algorithms can be modeled using classical Hamiltonians. That is not
surprising because the basic quantum equations (\ref{15}) are $c$-number
equations and can be formally considered as an effective classical
system of equations. The above results show that in some cases, effective
classical Hamiltonians can be used for simulation of quantum logic
operations even for large number of qubits. We present here the
corresponding effective classical Hamiltonian and the Hamiltonian
equations of motion in explicit form.

Formally, we can represent the solution of the Schr\"odinger equation
for the Hamiltonian (\ref{2})-(\ref{4}) in the form:
$\Psi(t)=\sum^{2^N-1}_{n=0}c_n(t)|n\rangle$, where $|n\rangle$ and
$E_n$ are the eigenfunctions and the eigenvalues of the Ising part
of the Hamiltonian in (\ref{2}) in the $z$-representation,
\begin{equation}
\label{43}
\hat {H}_0|n\rangle=E_n|n\rangle.
\end{equation}
The complex coefficients, $c_n(t)$, satisfy the equations,
\begin{equation}
\label{44}
i\dot c_n(t)=E_nc_n(t)+\sum_{k=0}^{2^N-1}{V}_{n,k}(t)c_k(t),
\end{equation}
where,
\begin{equation}
\label{45}
{V}_{n,k}=\langle n|{V}(t)|k\rangle,
\end{equation}
and ${V}(t)$ is the interaction Hamiltonian in (\ref{2}).

Equation (\ref{44}) can be written in Hamiltonian form:
\begin{equation}
\label{46}
i\dot c_n={{\delta H_{cl}}\over{\delta c^*_n}},\qquad
i\dot c^*_n=-{{\delta H_{cl}}\over{\delta c_n}},
\end{equation}
where $H_{cl}$ is the Hamiltonian of the equivalent ``classical'' system,
\begin{equation}
\label{47}
H_{cl}=\sum_{n=0}^{2^{N}-1}E_n|c_n|^2+\sum_{n,p=0}^{2^{N}-1}c_n^*V_{n,p}c_p.
\end{equation}
Instead of using complex ``classical'' variables, $c_n$ and $c^*_n$, one
can introduce the following real ``classical'' variables: a
``coordinate'', $x_n$, and a ``momentum'', $p_n$, using, for example,
the canonical transformation,
\begin{equation}
\label{48}
x_n={{1}\over{\sqrt{2}}}(c_n+c_n^*),\qquad
p_n=-{{i}\over{\sqrt{2}}}(c_n-c_n^*).
\end{equation}
Using $x_n$ and $p_n$, Eqs. (\ref{46}) take the familiar classical
Hamiltonian form,
\begin{equation}
\label{49}
\dot x_n={{\delta H_{cl}}\over{\delta p_n}},\qquad
\dot p_n=-{{\delta H_{cl}}\over{\delta x_n}},
\end{equation}
where the equivalent classical Hamiltonian is,
\begin{equation}
\label{50}
H_{cl}(x_n,p_n,t)=
\sum_{n=0}^{2^{N}-1}{{E_n}\over{2}}(x^2_n+p^2_n)+{{1}\over{2}}
\sum_{n,k=0}^{2^{N}-1}(x_n-ip_n)V_{n,k}(x_k+ip_k).
\end{equation}
The Hamiltonian (\ref{50}) can be written in the form,
\begin{equation}
\label{51}
H_{cl}(x_n,p_n,t)=\sum_{n=0}^{2^{N}-1}
{{E_n}\over{2}}(x^2_n+p^2_n)+
\end{equation}
$$
{{1}\over{2}}\sum_{n,k=0}^{2^{N}-1}
[x_n{\it Re}(V_{n,k})x_k+p_n{\it Re}(V_{n,k})p_k+2p_n{\it Im}(V_{n,k})x_k],
$$
where we used the relations: $V_{n,k}=V^*_{k,n}$, and
${\it Re}(V_{n,k})={\it Re}(V_{k,n})$,
${\it Im}(V_{n,k})=-{\it Im}(V_{k,n})$.
The corresponding classical Hamiltonian equations follow from
(\ref{49}) and (\ref{51}),
\begin{equation}
\label{52}
\dot x_n=E_nx_n+
\sum_{k=0}^{2^{N-1}}[{\it Re}(V_{n,k})p_k+2{\it Im}(V_{n,k})x_k],
\end{equation}
$$
\dot p_n=-E_np_n-
\sum_{k=0}^{2^{N-1}}[{\it Re}(V_{n,k})x_k-2{\it Im}(V_{n,k})p_k].
$$
For example, the Hamiltonian (\ref{51}) can be used to calculate the
dynamics of the quantum Hamiltonian (2). In this case, the matrix
elements, $V_{n,k}\not=0$ only for those states $|n\rangle$ which differ
from the state $|k\rangle$ by a single-spin transitions, and are zero
otherwise.

Solutions of Eqs. (\ref{51}) satisfy the normalization condition,
\begin{equation}
\label{53}
\sum_{n=0}^{2^{N}-1}[x^2_n(t)+p^2_n(t)]=1.
\end{equation}
Eqs. (\ref{51}) describe the Hamiltonian dynamics of $2^N$
classical ``generalized'' one-dimensional oscillators. Each of these
oscillators is described by two canonically conjugate variables, $x_n$
and $p_n$. One can see that $2^N$ classical harmonic oscillators can
simulate the behavior of $N$ interacting quantum qubits, including
the dynamics of quantum entanglement,
complicated quantum logic gates, and quantum algorithms.

\section*{Conclusion}
In this paper, we developed an approach, based on small parameters,
for the simulation of simple quantum logic operations in a nuclear spin
quantum computer with large number of qubits. We considered a quantum
computer which is a one-dimensional nuclear spin chain placed in a slightly
non-uniform magnetic field, oriented in a direction chosen to suppress the
dipole interaction between spins. We took into consideration the Ising
interaction between neighboring qubits. Quantum logic operations are
implemented by applying resonant electromagnetic pulses
to the nuclear spin chain. The electromagnetic pulse which is resonant
to a selected qubit is non-resonant for all other qubits. This raises
the problem of minimizing the influence of unwanted non-resonant effects
in the process of performing a quantum protocol.

We simulate the creation of the long-distance entanglement between remote
qubits, $(N-1)$-st and $0$-th, in a nuclear spin quantum computer having a
large number of qubits (up to 1000).
We used two essential assumptions:\\
1. The nuclear spin chain is prepared initially in its ground state. \\
2. The frequency difference between the neighboring spins due to the
inhomogeneity of the external magnetic field, $\delta\omega$, is much
larger than the Ising interaction constant, $J$,
and $J$ is much larger than the Rabi frequency, $\Omega$, i. e.
$\Omega\ll J\ll\delta\omega$.
Using these assumptions, we developed a numerical method which allowed us
to simulate the dynamics of quantum logic operations taking into
consideration all quantum states with the probability no less that
$P=10^{-6}$. For the case when the $2\pi k$ condition is satisfied
(the $\pi$-pulse for the resonant transition is at the same time a
$2\pi k$-pulse for non-resonant transitions), we obtained an analytic
solution for the evolution of the nuclear spin chain. In the case of
small deviations from the $2\pi k$ condition, the error accumulates and
the numerical simulations are
required. These results are presented in sections
\ref{sec:simulat} and \ref{sec:numerical}.

The main results of our simulations are the following:\\
1. The unwanted states exhibit a band structure in their probability
distributions. There are two main ``bands'' in the probability distribution
of unwanted states. The unwanted states in these ``bands'' have
significantly different probabilities, $P_{low}/P_{upper}\sim 10^{-3}$.
Each of these two bands have their own structures. \\
2. A typical unwanted state is a state of highly correlated spin
excitations. An important fact is that the unwanted states with
relatively {\it high
probability} include high energy states of the spin chain (many-spin
excitations).\\
3. The method developed allowed us to study the generation of unwanted
states and the probability of the desired states as a function of the
distortion of {\it rf} pulses. This can be used to formulate the
requirements for acceptable errors in quantum computation.

The results of this paper can be used to design experimental
implementations of quantum logic operations and to estimate (benchmark)
the reliability of experimental quantum computer devices. Our approach
can be extended to simulations of simple quantum arithmetic operations
and fragments of quantum algorithms. These simulations are now in
progress.

\end{document}